\documentclass{article}

\usepackage{arxiv}

\usepackage[utf8]{inputenc} 
\usepackage[T1]{fontenc}    
\usepackage{hyperref}       
\usepackage{url}            
\usepackage{booktabs}       
\usepackage{amsfonts}       
\usepackage{nicefrac}       
\usepackage{microtype}      
\usepackage{lipsum}		
\usepackage{tabularx}
\usepackage{multirow} 
\usepackage{adjustbox}
\usepackage{amssymb}
\usepackage{longtable}
\usepackage{booktabs}
\usepackage{pifont}
\usepackage{graphicx}
\usepackage[square,sort,comma,numbers]{natbib}
\usepackage{doi}
\usepackage{multirow}
\usepackage{float}
\usepackage{booktabs, makecell, tabularx}
\usepackage{graphicx}
\usepackage{booktabs}
\usepackage{amsmath}
\usepackage{multirow}
\usepackage{float}
\usepackage{graphicx}
\usepackage{xcolor}
\usepackage{hyperref}
\usepackage{multirow}
\usepackage{rotating}
\usepackage{pifont}

\usepackage{booktabs}
\usepackage{adjustbox}
\usepackage{tikz}
\newcommand{\redbullet}{\tikz[baseline=-0.5ex]\node[draw=red,fill=red,circle,inner sep=1.5pt]{};}
\newcommand*\greenbullet{\tikz[baseline=-0.75ex]\node[draw=green,fill=green,circle,inner sep=1.3pt]{};}

\title{Artificial Intelligence in Bone Metastasis Analysis: Current Advancements, Opportunities and Challenges}

\author{
\href{https://orcid.org/0000-0002-0255-7596}{\includegraphics[scale=0.06]{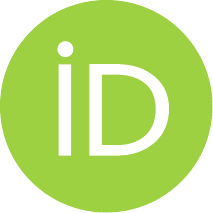}\hspace{1mm}Marwa AFNOUCH} \\
Universit{\'e} Polytechnique Hauts-de-France,\\ Université de Lille, CNRS, Valenciennes, \\  59313, Hauts-de-France, France\\
	\texttt{marwa.afnouch@uphf.fr} \\ 
 \And
 \href{https://orcid.org/0000-0001-5077-4862}{\includegraphics[scale=0.06]{orcid.pdf}\hspace{1mm}Fares BOUGOURZI}
 \\
Junia, UMR 8520, CNRS, Centrale Lille,\\ Univerity of Polytechnique Hauts-de-France,\\ 59000 Lille, France 
	\texttt{faresbougourzi@gmail.com} \\
	\And
	\href{https://orcid.org/0000-0002-1062-5528}{\includegraphics[scale=0.06]{orcid.pdf}\hspace{1mm}Olfa Gaddour} \\
	National engineering school of Sfax, Tunisia\\
	\texttt{olfa.gaddour@enis.tn} \\ 
	\And
	\href{https://orcid.org/0000-0001-6581-9680}{\includegraphics[scale=0.06]{orcid.pdf}\hspace{1mm}Fadi DORNAIKA} \\
	University of the Basque Country UPV/EHU,\\
	San Sebastian, SPAIN; IKERBASQUE, Basque \\ Foundation for Science, Bilbao, SPAIN \\
	\texttt{fadi.dornaika@ehu.eus} \\ 
	\And
	\href{https://orcid.org/0000-0001-7218-3799}{\includegraphics[scale=0.06]{orcid.pdf}\hspace{1mm}Abdelmalik Taleb-Ahmed} \\
	Universit{\'e} Polytechnique Hauts-de-France, Université de Lille, \\CNRS, Valenciennes, 59313, Hauts-de-France, France\\
	\texttt{Abdelmalik.Taleb-Ahmed@uphf.fr} \\
}

\renewcommand{\shorttitle}{\textit{arXiv} Template}

\hypersetup{
pdftitle={A template for the arxiv style},
pdfsubject={q-bio.NC, q-bio.QM},
pdfauthor={David S.~Hippocampus, Elias D.~Striatum},
pdfkeywords={First keyword, Second keyword, More},
}

\begin{document}
\maketitle

\begin{abstract}
In recent years, Artificial Intelligence (AI) has been widely used in medicine, particularly in the analysis of medical imaging, which has been driven by advances in computer vision and deep learning methods. This is particularly important in overcoming the challenges posed by diseases such as Bone Metastases (BM), a common and complex malignancy of the bones. Indeed, there have been an increasing interest in developing Machine Learning (ML) techniques into oncologic imaging for BM analysis.
In order to provide a comprehensive overview of the current state-of-the-art and advancements for BM analysis using artificial intelligence, this review is conducted with the accordance with PRISMA guidelines.
Firstly, this review highlights the clinical and oncologic perspectives of BM  and the used medical imaging modalities, with discussing their advantages and limitations.
Then the review focuses on modern approaches with considering the main BM analysis tasks, which includes:  classification, detection and segmentation. 
The results analysis show that ML technologies can achieve promising performance for  BM analysis and have significant potential to improve clinician efficiency and cope with time and cost limitations. Furthermore, there are requirements for further research to validate the clinical performance of ML tools and facilitate their integration into routine clinical practice.
\end{abstract}

\keywords{Bone Metastases \and Dataset \and Machine Learning \and Deep Learning \and CNN \and Artificial Intelligence \and Medical Imaging Analysis \and Transformer \and CNN \and GAN}

\section{Introduction}
Metastasis is the process by which an accumulation of abnormal cells proliferates beyond the limits of the primary organ and spreads to distant organs of the body. The bones are one of the most common sites for these metastases, third only to the lungs and liver \cite{tsuya2007skeletal}. Breast and prostate cancer are the main causes of primary tumors and account for about 70\% of cancer cases \cite{siegel2022cancer}. In most cases, complete eradication of bone lesions is rare and patients usually undergo palliative treatment to shrink the lesion, slow growth or improve symptoms. The average survival time of bone metastases (BM) in breast cancer is 19-25 months and in prostate cancer 53 months, resulting in a significant reduction in life expectancy. In addition, BM can cause complications such as pain due to fractures, reduced mobility and neurological deficits. They can also lead to symptoms such as constipation, fatigue, cardiac arrhythmias and acute renal failure \cite{coleman2001metastatic}. Early detection, diagnosis and appropriate treatment of bone metastases is therefore essential to reduce complications and improve patients’ quality of life.

The rapid advances in the field of artificial intelligence (AI), particularly in the areas of big data management and image processing, have generated much optimism for its integration into the medical field, especially in medical imaging \citep{anthimopoulos2016lung,bougourzi2023pdatt}. In particular, machine learning, especially deep learning, has made remarkable progress in medical image analysis, encompassing tasks such as image recognition and bioinformatics.
The growing volume of biomedical data, including the exponential growth of medical images enabled by high-throughput technologies developed over the past decades \citep{litjens2017survey,bougourzi2024emb}, has found an ideal companion in AI systems. These systems have demonstrated a remarkable ability to automatically recognize nuanced details in medical images.

In the context of BM, AI enables quantitative assessments that differ from the subjective visual assessments of clinicians. In addition, AI systems show promise in addressing potential shortcomings of human expert diagnoses, such as the tendency to overlook small metastatic lesions, thus reducing the risk of misdiagnosis \citep{hosny2018artificial}.
As the development of efficient AI systems is crucial for BM analysis, conducting a comprehensive survey in this area is of paramount importance. The main goal is to facilitate access to and use of resources by researchers from diverse backgrounds to drive advances in BM research. This includes a thorough discussion and analysis of the current state of the art and results related to various BM analysis tasks, including classification, segmentation, detection, and registration.

In order to achieve this goal, the most significant works on BM analysis from the last thirteen years have been gathered. The methods and results presented in the individual articles are explained and discussed in detail. The collected papers have been categorized into four main groups for BM analysis, which include classification, segmentation, detection, and other ML tasks such as registration. The literature within these categories is further subdivided based on the different medical imaging modalities.
This review not only highlights and discusses the most compelling techniques in the literature, but also identifies and discusses research gaps that require further investigation and outlines future research directions. Moreover, It serves to establish a link between the computer vision community and the medical community to promote collaboration and support future research and development in the field of bone metastasis analysis. This survey provides the following contributions::

\begin{itemize}
\item An examination of recent surveys on interpretable ML in BM analysis, including the main conclusions derived from each, as well as a comparison with our research. 

\item A list of all the medical imaging modalities commonly utilized in the study of interpretability of ML approaches applied to BM analysis.

\item A detailed description of the datasets typically utilized to evaluate ML methods for BM analysis.

\item A comprehensive review of the current state-of-the-art ML and DL methods employed in BM analysis. Our analysis goes beyond a mere summary of these methods and includes an in-depth examination of their strengths and limitations.
\item Research directions in interpretable ML in BM analysis in the future.

\end{itemize}

The subsequent sections of this review are structured as follows. Section \ref{sec2} outlines the meticulous methodology employed for selecting and evaluating relevant literature. Section \ref{sec3} provides a comprehensive overview of bone metastasis, its clinical significance, and current diagnostic approaches. Section \ref{sec4} then delves into the various imaging modalities utilized in bone metastasis detection, highlighting their strengths and limitations. Section \ref{sec5} critically assesses the characteristics of publicly available datasets commonly used for evaluating ML and DL models in this field. Section \ref{sec6} explores the diverse ML tasks involved in bone metastasis analysis, ranging from lesion detection and segmentation to risk prediction and treatment response assessment. Section \ref{sec8} engages in a critical discussion of the challenges and promising future directions for research in this domain, considering both technical and clinical perspectives. Finally, Section \ref{sec9} provides a concise and insightful conclusion that summarizes the key findings and reiterates the potential impact of ML and DL in advancing bone metastasis analysis.

\section{Survey Methodology}
\label{sec2}

\subsection{Related Surveys}
\begin{table}[htbp]
\small
\centering
\caption{Comparative Analysis between the Surveys on the Topic of Artificial Intelligence Applied to Bone Metastasis Imaging Analysis. "BS=Bone Scintigraphy, NSB= None Spinal Bone (denotes surveys encompassing papers discussing BM in all regions except the spinal region), SB=Spinal Bone (refers to surveys focusing exclusively on papers discussing proposed solutions for spinal metastasis), Hybrid=Hybrid imaging modalities like SPECT/CT or PET/CT, CT=Compured Tomography, MRI= Magnetic Resonance Imaging, SPECT=Single Photon Emission Computed Tomography, BM regions= Bone Metastasis location in the Human body. Red bullets indicate not available, and green bullets indicate available."}
\label{tab:comparative}
\adjustbox{width=\textwidth}{\begin{tabular}{llccccccccccccc}
\toprule
Ref. & Year & \begin{tabular}[c]{@{}l@{}}Dataset \\Description\end{tabular} & \multicolumn{5}{c}{Imaging Modalities} & \multicolumn{2}{c}{BM Regions} & \multicolumn{4}{c}{Analysis Tasks}\\
\cmidrule(lr){4-8} \cmidrule(lr){9-10} \cmidrule(lr){11-14}
 & & & BS & MRI & CT & SPECT & Hybrid & TB & SB & Classification & Segmentation & Detection & Others \\
\midrule

\cite{chen2022application} & 2022 & \redbullet  & \redbullet & \redbullet  & \greenbullet   & \greenbullet & \greenbullet   & \greenbullet&\redbullet & \greenbullet& \greenbullet& \greenbullet &\redbullet\\
\cite{Zhou_2022} & 2022  &\redbullet& \greenbullet  & \greenbullet  & \greenbullet  & \redbullet & \redbullet & \greenbullet & \redbullet & \greenbullet  & \redbullet&\redbullet& \redbullet   \\
\cite{faiella2022artificial} & 2022 & \redbullet & \redbullet & \greenbullet & \greenbullet & \redbullet& \redbullet& \greenbullet & \greenbullet & \greenbullet &\greenbullet & \redbullet &\redbullet \\
\cite{ong2022application} & 2022 & \redbullet & \redbullet& \greenbullet & \greenbullet & \redbullet& \redbullet & \redbullet  & \greenbullet & \greenbullet & \greenbullet& \greenbullet &\redbullet \\
\cite{kao2023systematic} & 2023 & \redbullet& \greenbullet& \redbullet& \redbullet & \redbullet & \redbullet &\greenbullet  &\redbullet & \greenbullet & \redbullet & \redbullet  &\redbullet \\
\cite{paranavithana2023systematic} & 2023& \redbullet & \greenbullet & \redbullet & \greenbullet& \redbullet & \greenbullet & \greenbullet & \greenbullet & \redbullet& \greenbullet & \redbullet &\redbullet \\
\cite{kendrick2021radiomics} & 2023& \redbullet &\redbullet & \greenbullet & \greenbullet & \redbullet& \greenbullet& \greenbullet & \redbullet& \greenbullet & \redbullet & \greenbullet  & \redbullet \\
\textbf{Ours}& 2024 & \greenbullet& \greenbullet& \greenbullet & \greenbullet & \greenbullet & \greenbullet & \greenbullet & \greenbullet & \greenbullet & \greenbullet & \greenbullet  &\greenbullet \\
\bottomrule
\end{tabular}}
\label{tabsurvey}
\end{table}

In recent years, there has been increasing interest in Bone Metastasis (BM) analysis, leading to a rise in the number of reviews. Despite the observed increase, many aspects have not been adequately addressed or have been entirely overlooked. Table \ref{tabsurvey} presents a comparative analysis between existing surveys and ours. As shown, many existing surveys have not covered important aspects of BM analysis.

A significant limitation of existing reviews is the lack of description of the datasets used, and they have not discussed the availability of the data, which plays a crucial role in advancing BM analysis. Moreover, many medical imaging modalities have not been addressed in previous surveys. While existing surveys addressed up to three modalities, our survey addresses four medical imaging modalities plus the hybrid one. Existing surveys primarily focused on specific BM regions, particularly the spine, while other critical metastatic sites such as the pelvis, femur, and foot were often overlooked \cite{ong2022application}. It is essential to note that the spine is the most common site for BM. Conversely, some surveys have focused exclusively on the entire bone structure and have disregarded papers specifically addressing spinal metastases \cite{chen2022application,Zhou_2022}, despite their clinical significance. This exclusion limits the comprehensive understanding of BM analysis.

For the analyzed tasks, some existing surveys concentrated only on one task, such as classification \cite{kao2023systematic} or segmentation \cite{paranavithana2023systematic}. On the other hand, other surveys focused on only two tasks \cite{faiella2022artificial,kendrick2021radiomics}. Only two out of seven state-of-the-art surveys covered the three essential BM analysis tasks (classification, segmentation, and detection) \cite{chen2022application,ong2022application}. However, these surveys discarded other tasks, including registration and multi-tasks, which highlight important aspects in BM analysis.

Moreover, Table \ref{compsur2} depicts a comparison with state-of-the-art surveys in terms of the number of papers, time coverage, machine learning method, training paradigm, and deep learning architecture. Similar to observations from Table \ref{tabsurvey}, the comparison shows many limitations in previous surveys that need further investigation, including the limited number of papers addressed and the time coverage, with the most recent paper being from 2022 in previous surveys and 2024 for our survey. Despite most previous surveys discussing both Deep and shallow methods, the investigation of training paradigms was limited to the Supervised manner. In contrast, we cover other training paradigms (Semi-supervised and Unsupervised) for their high capability to cope with data labelling limitations, which are the main challenging aspects toward developing efficient real-time BM analysis. Moreover, previous surveys concentrated only on Convolutional Neural Networks (CNNs) from deep learning approaches, since Transformer is a new approach and the previous surveys' time limitation is 2022. Transformers have promising performance for vision tasks, including BM analysis. These limitations highlight the high need for an overall comprehensive survey in the field of BM analysis to help understand the current state-of-the-art, limitations, and the required steps to reach the objective of efficient real-world applications.

Our survey aims to thoroughly explore how AI can be used in BM analysis, covering various aspects of this field. To achieve this, we have outlined the different ways AI can be applied in BM analysis and given a detailed overview of the diagnostic process. By analyzing, summarizing, and comparing current Machine Learning (ML) and Deep Learning (DL) models, this review aims to provide a comprehensive understanding of the field and inspire future research. What sets us apart is our focus on the imaging methods used, along with detailed insights into model designs, databases, and practical uses. Additionally, our study stands out for directly tackling the unique challenges of BM analysis. This not only helps advance the field but also drives progress in diagnosing and treating BM-related conditions.

\begin{table}[ht]
\caption{Comparison of State-of-the-Art Surveys: Number of Papers, Time Coverage, Machine Learning Method, Training Paradigm, and Deep Learning Architecture.}
\adjustbox{width=\textwidth}{
\begin{tabular}{l|c|c|cc|ccc|cc}
\hline
\multirow{2}{*}{Ref}  & \multirow{2}{*}{Num.papers} & \multirow{2}{*}{Coverage}  & \multicolumn{2}{c|}{ML Method} & \multicolumn{3}{c|}{Training Paradigm}             & \multicolumn{2}{c}{DL Arch Type}                              \\ \cline{4-10} 
   & & & \multicolumn{1}{c}{DL methods} & Shallow methods & \multicolumn{1}{c}{Supervised} & \multicolumn{1}{c}{Semi-supervised} & Unsupervised & \multicolumn{1}{c}{CNN} & \multicolumn{1}{c}{Transformer}  \\ \hline
\cite{chen2022application}   & 25   & N/A & \greenbullet  & \redbullet   & \greenbullet  & \redbullet   & \redbullet & \greenbullet   & \redbullet            \\ 
\cite{Zhou_2022}  & 22 & 2015-2021   & \greenbullet  & \redbullet         & \greenbullet& \redbullet    & \redbullet    & \greenbullet   & \redbullet                \\ 
\cite{faiella2022artificial} & 13   & 2018-2021 & \greenbullet  & \greenbullet  & \greenbullet  & \redbullet   & \redbullet    & \greenbullet   & \redbullet     \\ 
\cite{ong2022application}   & 30    & N/A   & \greenbullet & \greenbullet& \greenbullet & \redbullet   & \redbullet    & \greenbullet   & \redbullet   \\ 
\cite{kao2023systematic}  & 24  & N/A                                 & \greenbullet  & \greenbullet & \greenbullet & \redbullet    & \redbullet    & \greenbullet   & \redbullet     \\ 
\cite{paranavithana2023systematic} & 77                                & 2010-2022 & \greenbullet   & \greenbullet & \greenbullet  & \greenbullet  & \redbullet   & \greenbullet   & \redbullet      \\ 
\cite{kendrick2021radiomics}  & 30   & 2015-2021  & \greenbullet          & \greenbullet  & \greenbullet  & \redbullet  & \redbullet   & \greenbullet & \redbullet      \\ 
\textbf{Ours}   & \textbf{95}   & \textbf{2012-2024} & \greenbullet    & \greenbullet  & \greenbullet  & \greenbullet   & \greenbullet & \greenbullet   & \greenbullet     \\ \hline
\end{tabular}
}
\label{compsur2}
\end{table}


\begin{figure}[htbp]
\centering
\includegraphics[width=1.\textwidth]{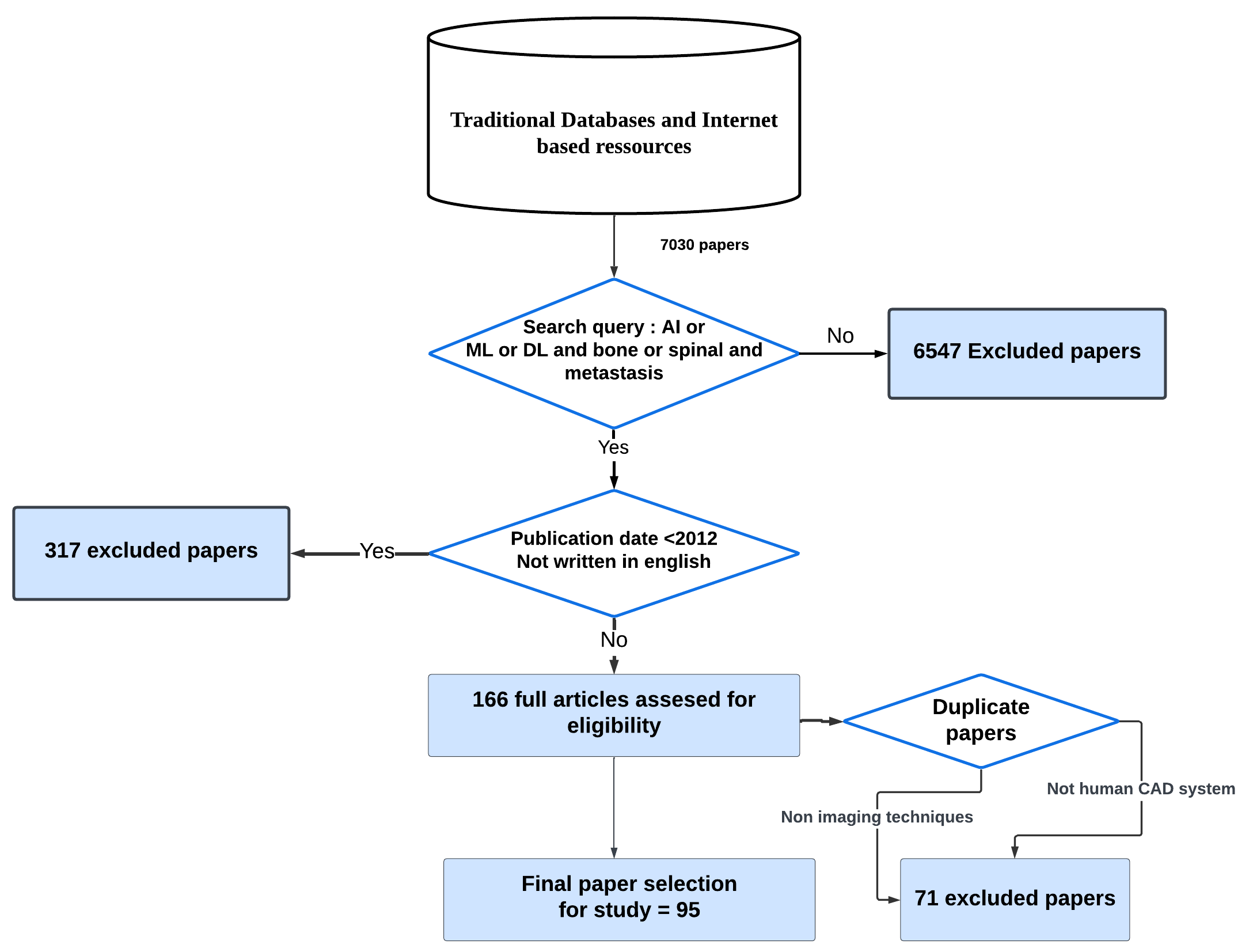}
\caption{Survey Search Strategy }
\label{fig1}
\end{figure}

\subsection{Papers Selection}

In recent years, AI for BM analysis has witnessed a surge of research in well-known AI and electronic medical journal databases including Web of Science, clinicaltrials.gov, MEDLINE, and PubMed. Figure \ref{fig1} summarizes the search strategy used in this study. The search focuses on finding related papers on AI for BM analysis and its applications in recent years. To this end, the search has been limited to articles published up to January 31, 2024, and a set of predefined terms is used: 'Artificial intelligence', 'radiomics' or 'machine learning' or 'deep learning' and 'bone', 'spinal' or 'skeletal' and 'metastasis' or 'metastases' or 'metastatic.' AI for BM includes many subareas such as image analysis, prognosis prediction, treatment prediction, etc. Moreover, additional keywords related to the sub-fields of AI for BM analysis are used to get more relevant articles, such as "Classification", "Segmentation", "Machine Learning", and so on. 

In the literature search, no limitations were specified. An important inclusion criterion was the use of AI techniques to analyze BM disease. Among all studies collected from electronic medical databases, the following criteria were used to identify convenient studies that met the requirement for the research question. 
\begin{itemize}
\item The work should be published between 2012 and 2024.
\item The study must be written in English and involve only humans.
 \item The article found must address a sub-direction of BM or the important axis of BM evolution in terms of analysis, medical imaging and downstream task evaluation along with the commonly used datasets.
\end{itemize}
For the purpose of the analysis, certain types of articles were excluded. This included case reports, abstracts from conferences, review articles, and editorial correspondence such as letters, commentaries, and opinion pieces. Additionally, duplicate publications and publications that only focused on nonimaging radiomic techniques, specifically those that only analyzed isolated histopathology features, were also excluded from the analysis. To ensure that all relevant articles were included, the final step of the literature search involved a manual review of the bibliographies of the selected publications. 

\subsection{Papers Distribution}
\begin{figure}[htbp]
\centering
\includegraphics[width=0.9\textwidth]{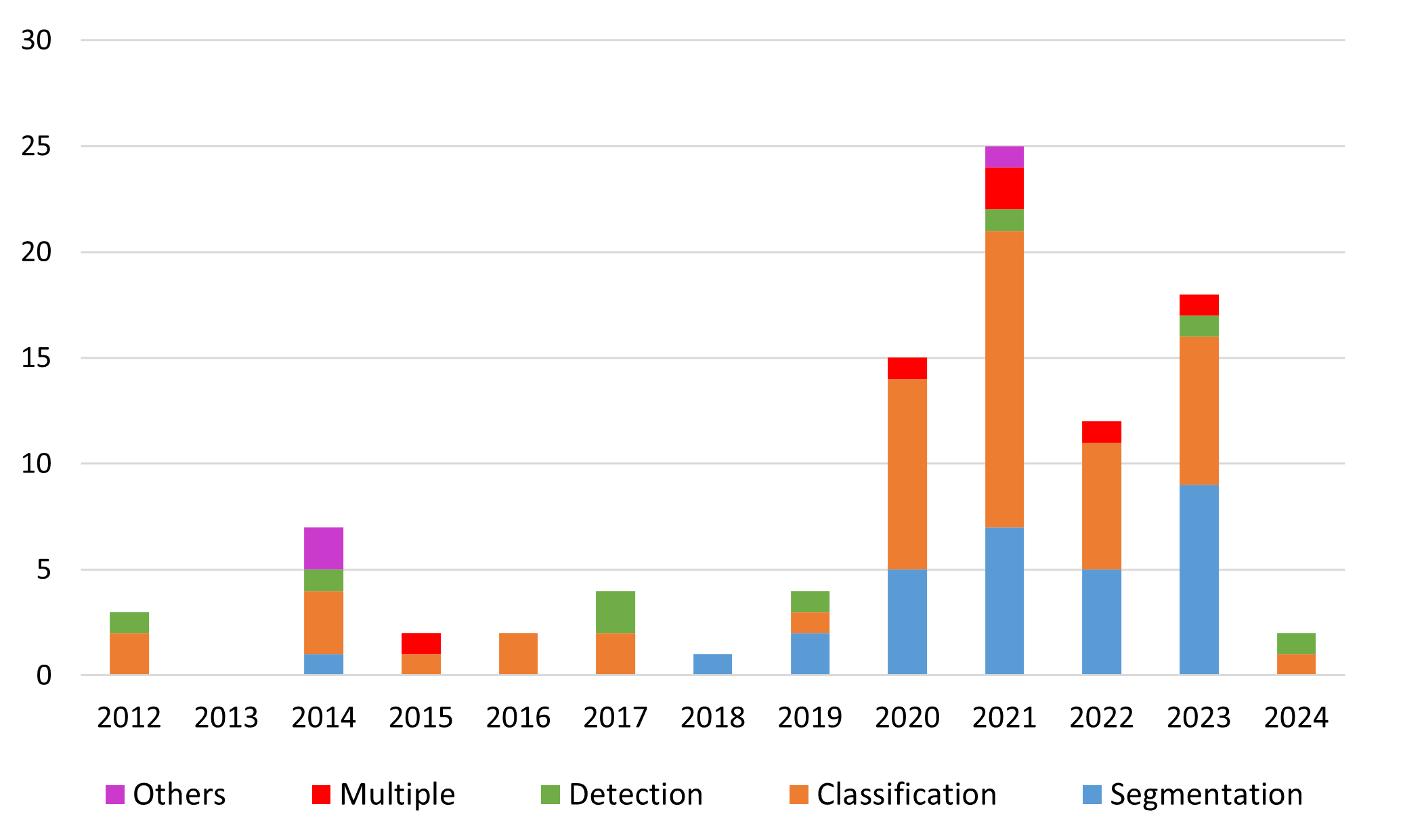}
\caption{A chronological distribution of artificial intelligence research publications in bone metastasis analytic }
\label{year}
\end{figure}

Figure \ref{year} shows a visually informative histogram depicting the distribution of reviewed articles over the different publication years. The color distinction facilitates the identification of the different ML tasks studied in the literature, including classification, segmentation, detection, multiple, and other tasks. The histogram shows a notable trend in the application of AI techniques for BM analysis, with interest increasing significantly from 2020 onward. The year 2021 is particularly striking, with more than 23 articles published, indicating a significant increase in research and exploration in this area. These results underline the growing importance of computer vision and AI methods in addressing the challenges of BM analysis. While the graph provides an overview of the distribution of papers over the years and the extensive range of ML tasks investigated, a more comprehensive analysis of the specific trends, methods and results is presented in section \ref{sec6} of the paper.

\section{Bone Metastasis: Overview}
\label{sec3}
\subsection{Bone Metastases in Medical Oncology}

Bone metastases, or bone mets or bone metastatic disease, are a medical condition in which cancer cells from a primary tumor spread to the bone. Although they can occur in any type of cancer, they are commonly seen in breast and prostate cancers \cite{Macedo_2017}. Invasion of cancer cells into bones can cause significant damage and fragility, resulting in severe pain, fractures, and other complications. Furthermore, cancer cells can disrupt normal bone function, resulting in decreased bone strength and an impaired ability to heal itself. In addition, BM can attack the nerves and spinal cord, causing numbness, tingling and even paralysis \cite{Shupp_2018}. This is a serious complication of cancer that is a major health problem and can be difficult to manage. BM is a major cause of morbidity in people with cancer \cite{Macedo_2017}.
\begin{figure}[h]
\centering
\includegraphics[scale=0.6]{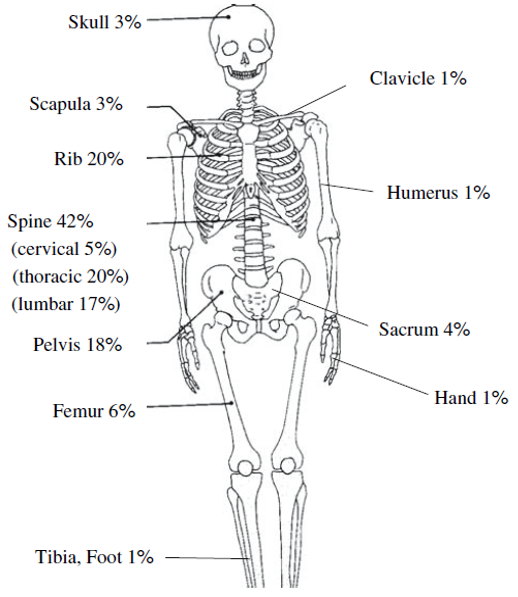}
\caption{Anatomic localization of skeletal metastases from lung cancer \cite{sugiura2008predictors} }
\label{fig11}
\end{figure}
Skeletal metastases tend to occur in body regions where red bone marrow is most abundant, suggesting that the greater blood supply to red bone marrow compared with yellow bone marrow is an important factor. The most commonly affected sites include the vertebrae, favoring the lumbar spine over the thoracic and cervical spine, as well as the pelvis, proximal femur, ribs, and scapula, as shown in Figure \ref{fig11}. On the other hand, metastases to distal limb sites, such as the elbow and knee, are rare. In summary, the distribution of BM in the body is not random and tends to occur in areas with a high concentration of red bone marrow. This underscores the importance of understanding the underlying mechanisms that drive the development and spread of cancer cells in these areas \cite{kakhki2013pattern}.

\begin{figure}[htbp]
\centering
\includegraphics[width=0.9\textwidth]{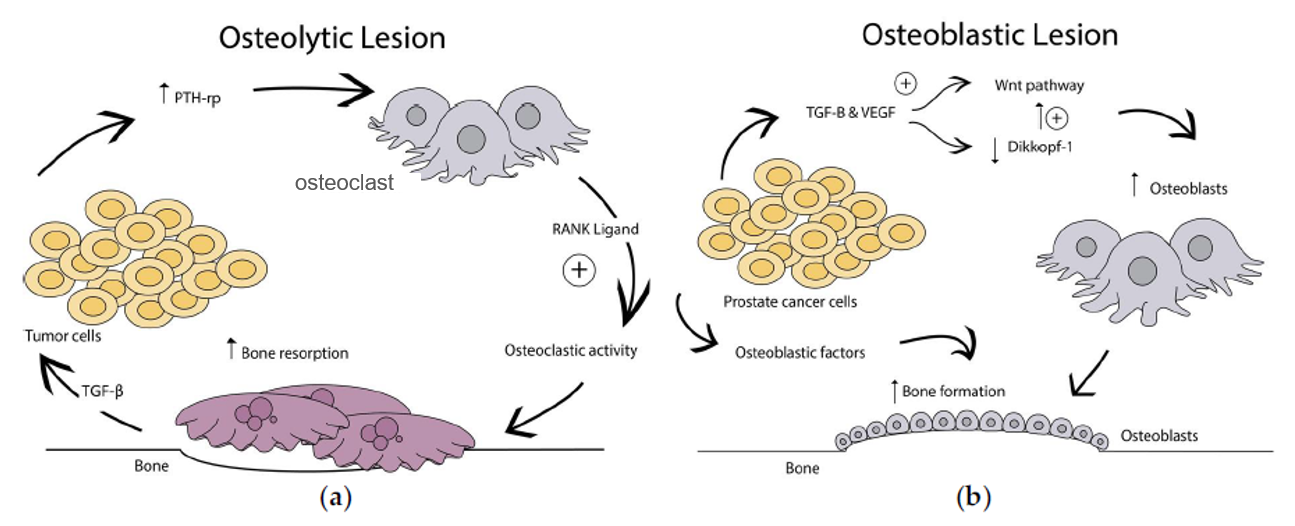}
\caption{Metastatic Bone Lesions \cite{Jinnah_2018} }
\label{fig3}
\end{figure}

BM may occur through hematogenous or lymphatic routes or by direct spread of tumors to the bone. Regardless of the route of spread, BM results in bone loss and bone formation, leading to different patterns of bone destruction and remodeling \cite{paduch2016role}. BM can adopt one of three dominant patterns, including lytic (osteolytic), sclerotic (osteoblastic), or mixed lytic and sclerotic metastases. Figure \ref{fig3} illustrates lytic and sclerotic bone lesions. In addition, BM may also have different morphological features, including diffuse, focal, or expansive. Although lytic metastases cause bone destruction, sclerotic metastases promote bone formation \cite{Jinnah_2018}.

Diagnosis of BM can sometimes be difficult, especially in elderly patients with degenerative diseases and osteoporosis. Imaging tests and serum tumor markers are critical for diagnosis; In some cases, a bone biopsy may be required. Treatment of BM is aimed at relieving symptoms, as a cure is rarely possible. Treatment options include external beam radiation therapy, endocrine treatments, chemotherapy, targeted therapies, radioisotopes, and orthopedic surgery for structural complications. The choice of treatment depends on the extent of bone disease, the presence of metastases outside the skeleton, and the underlying malignancy. As the disease progresses, resistance to systemic treatments can develop, requiring a change in therapy. 

\subsection{Common Primary Cancers Leading to Bone Metastases}

BM happens when cancer cells migrate to the bones from other parts of the body. According to \cite{masada2023fixation}, breast, prostate, and lung cancers account for about 80\% of all primary cancers that metastasize to the skeleton. In addition, bone metastases from thyroid and kidney cancers are also possible. Bone is the third most common site of metastatic disease after the lung and liver. According to the study in \cite{macedo2017bone} , breast and prostate cancers can cause up to 70\% of skeletal metastases. Lung cancer, renal cell carcinoma, and prostate cancer are the top three cancers that most commonly cause bone metastases. When the tumor and microenvironment are susceptible, bone metastases occur more frequently. Skeletal problems associated with advanced cancer are common in patients \cite{von2017improving}. 
In fact, the presence of bone metastases significantly shortens patient survival and makes it challenging to live beyond a relatively short period of time, as seen in Table \ref{tab2}.

\begin{table}
\centering
\caption{Incidence of BM in cancer (Data in the table based on  \cite{macedo2017bone} and \cite{suva2011bone}) }
\label{tab2}
\begin{tabular}{lllllll}
\cline{1-3}
Primary cancer type & Relative incidence in bone & Median survival from diagnosis &  &  &  &  \\ \cline{1-3}
Breast              & 65\% – 75\%                  & 19 – 25   months                     &  &  &  &  \\
Prostate            & 65\% – 75\%                  & 12 – 53   months                     &  &  &  &  \\
Lung                & 30\% – 40\%                  & 6   months                         &  &  &  &  \\
Thyroid             & 40\% – 60\%                  & 48   months                        &  &  &  &  \\
Bladder             & 40\%                       & 6–9   months                       &  &  &  &  \\
Renal               & 20\% – 25\%                  & 12   months                        &  &  &  &  \\
Melanoma            & 14\% – 45\%                  & 6   months                         &  &  &  &  \\ \cline{1-3}
                    &                            &                                &  &  &  & 
\end{tabular}
\end{table}

\section{Imaging Modalities for Bone Metastasis Diagnosis}
\label{sec4}
Imaging is essential to treating bone lesions \cite{rybak2001radiological}. Various morphological and functional imaging techniques are utilized to assess malignant bone involvement. The most commonly used imaging modalities include positron emission tomography (PET), computed tomography (CT), magnetic resonance imaging (MRI), single photon emission computed tomography (SPECT), and bone scintigraphy (BS) \cite{dotan2008bone}. These imaging modalities have proven important in detecting, segmenting, and diagnosing BM. It is essential to note that the choice of imaging modality depends on various factors, including the primary cancer site, the suspected extent of metastatic disease, and the specific clinical scenario.
\begin{itemize}
    \item Bone Scintigraphy, also called Bone Scan, is an imaging modality that involves injecting a radioactive substance into a patient's body. It is used to examine the various bones of the skeleton. Two gamma cameras are placed in front and behind the patient to detect radiation emitted from the injected radioactive substance. The two resulting images will show hotspots with high intensities. As the bone scan is widely available and less expensive, it is the most widely used radionuclide technique to evaluate skeletal metastases \cite{orcajo2022review}.
    \item CT scans use X-rays to produce detailed cross-sectional images of the body. CT imaging provides high-resolution anatomical information, allowing for precise localization and characterization of bone lesions \cite{O_Sullivan_2015,bauerle2009imaging}.
    \item MRI is a radiation-free, noninvasive imaging modality that uses a magnetic field and radio waves to create precise images of the inside of the body and provide 3D visualization of tissues \cite{florkow2022magnetic}.
    \item SPECT is one of the most commonly used techniques that uses trace concentrations of radioactively labeled compounds to provide insight into physiological processes. In SPECT examination, imaging equipment captures the emitted gamma rays from radionuclides that were injected into a patient’s body in advance to generate a map of the inside of a body \cite{heindel2014diagnostic}.
    \item PET is a nuclear medicine imaging technique that produces 3D anatomical information or maps of functional processes in the body. A small amount of radioactive material is injected into the patient's bloodstream and the gamma rays emitted by the radioactive material are detected by the PET scanner. Because bone metastases often have increased blood flow, PET can be used to detect the abnormal metabolic activity of cancer cells at multiple sites throughout the body in a single scan. The applicability and usefulness of PET images depends, among other things, on the radiopharmaceutical used. However, PET images have a higher resolution compared to conventional planar and SPECT techniques \cite{orcajo2022review}.
    \item Hybrid imaging has an increasing role in the early detection of BM and in monitoring response at early time points \cite{cook2020molecular}. In this sense, functional imaging such as SPECT/CT, PET/CT, and PET/MRI provides a standardized uptake value and allow the fusion of anatomic data from cross-sectional imaging with functional information from nuclear medicine studies. As a result, the radiologist can determine if focal radiotracer uptake on a nuclear medicine study corresponds to a discrete bone lesion.  Similarly, diagnostic confidence increases when an osseous lesion suspicious for metastasis on cross-sectional imaging avidly accumulates radiotracer. 
\end{itemize}

Each imaging modality has its strengths and weaknesses, as shown in Table \ref{tabimg}, which influence their utility in specific clinical scenarios. Combining different modalities can increase sensitivity and specificity, leading to improved diagnostic accuracy \cite{yang2011diagnosis}. Therefore, a multidisciplinary approach involving radiologists, oncologists, and other healthcare professionals is crucial in determining each patient's most appropriate imaging strategy.

\begin{table}[htbp!]
  \centering
  \caption{Comparative Analysis of Imaging Modalities for the Assessment of Bone Metastasis: Advantages and Limitations}
  \label{tab:imaging_modalities}
  \begin{tabular}{p{3cm}p{6cm}p{5cm}}
    \toprule
    \textbf{Imaging Modality} & \textbf{Advantages} & \textbf{Limitations} \\
    \midrule
    Bone Scan & 
    Sensitive to early bone changes \newline Whole-body coverage \newline Detects multiple lesions simultaneously &
    Low specificity \newline Cannot provide precise anatomical detail \newline Requires radiotracer injection \\
    \addlinespace
    \hline
    CT Scan &
    High spatial resolution \newline Detailed visualization of bone structures \newline Rapid scanning &
    Ionizing radiation exposure \newline Limited soft tissue contrast \newline May miss small lesions \\
    \addlinespace
    \hline
    MRI &
    Excellent soft tissue contrast \newline Multiplanar imaging capability \newline No ionizing radiation exposure &
    Expensive and time-consuming \newline Not widely available \newline Motion artifacts may affect images \\
   \addlinespace
    \hline
    SPECT Scan &
    Three-dimensional functional information \newline Assess bone metabolism and blood flow &
    Lower spatial resolution compared to PET \newline Longer acquisition times \\
    \addlinespace
    \hline
    PET Scan &
    Provides functional information about metabolic activity \newline Particularly useful in identifying sites of active tumor growth &
    Requires cyclotron for radiotracer production \newline Higher cost compared to other modalities \\
    \addlinespace
    \hline
    Hybrid Methods &
    Combined advantages of different modalities \newline Improved sensitivity and specificity \newline Enhanced anatomical and functional data &
    Costly \newline Limited availability \newline Longer scan times \\
    \bottomrule
  \end{tabular}
  \label{tabimg}
\end{table}

\begin{figure}[h]
\centering
\includegraphics[width=0.8\textwidth]{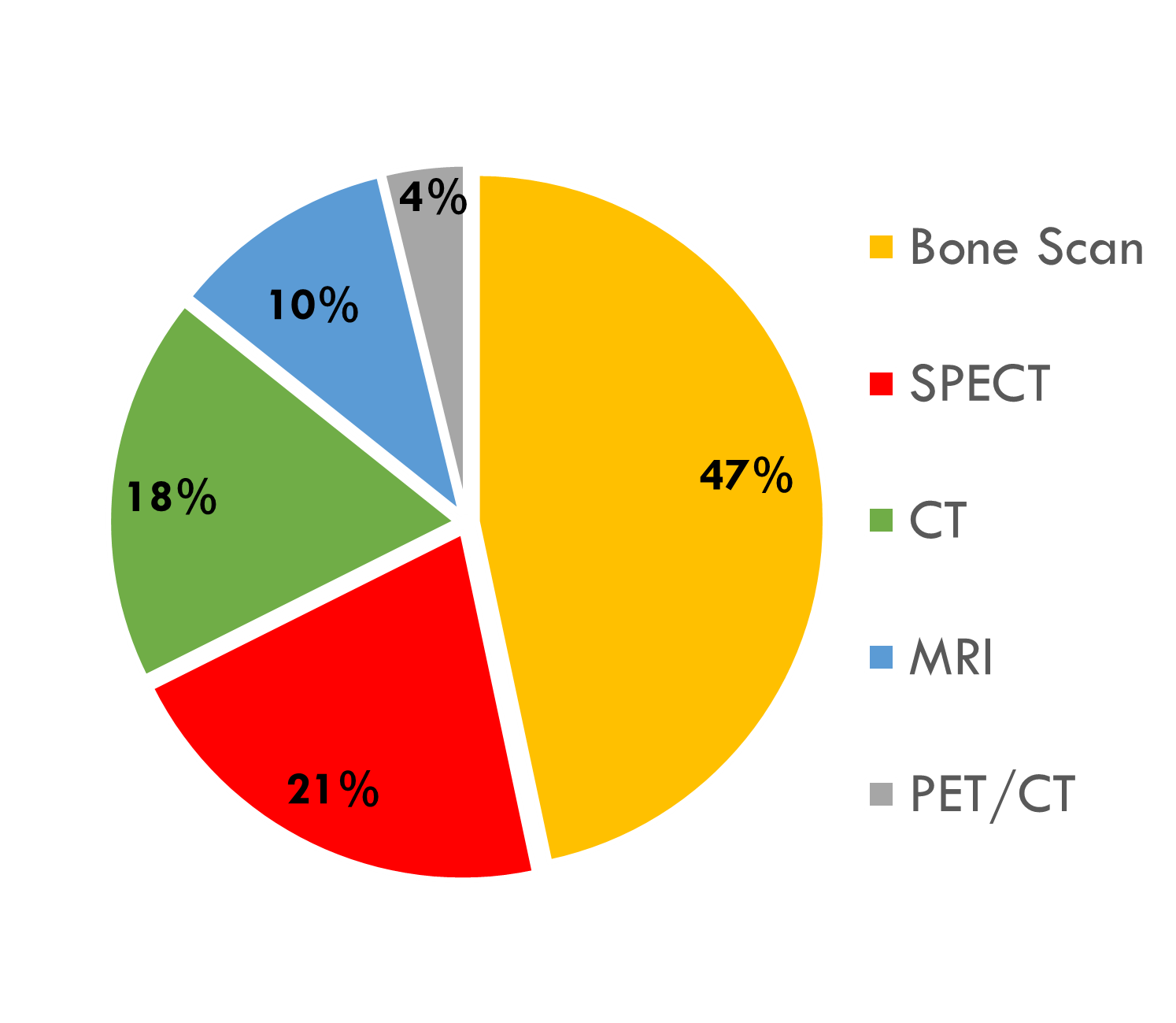}
\caption{Distribution of Medical Imaging Modalities in Reviewed Articles }
\label{modality}
\end{figure}

In addition, utilization rates of these modalities can vary due to factors such as accessibility, expertise and local practices. Figure \ref{modality} shows the distribution of imaging modalities used in the research papers. It reveals that bone scintigraphy is the most commonly used imaging modality, appearing in 47\% of the papers. This observation underscores the continued importance and widespread use of bone scans in the clinical setting for the detection and evaluation of bone metastases.

After bone scintigraphy, SPECT is the most frequently used imaging modality, accounting for 21\%. This indicates that SPECT imaging has established itself as an important tool for the assessment of bone metabolism and blood flow and provides valuable functional information in the context of BM. Figure \ref{modality} also shows that CT scans, including spectral energy CT scans, account for a significant proportion of utilization at 18\%. This observation is consistent with the known advantages of CT imaging, such as high spatial resolution and detailed visualization of bone structures. The inclusion of spectral-energy CT scans is another indication of the recognition of advanced CT techniques to improve diagnostic capabilities. MRI follows with a utilization rate of 10\%. This result suggests that although MRI provides excellent soft tissue contrast and multiplanar imaging, it is less utilized compared to other modalities such as bone and CT scans. Factors such as cost, availability, and the specific clinical context may contribute to this difference in utilization. Finally, the hybrid imaging modality PET/CT is also used to a lesser extent, in only 4\% of papers included in this survey. Although PET/CT combines the advantages of PET and CT and provides functional and anatomical information, its relatively low utilization rate can be attributed to factors such as higher costs and the need for dedicated infrastructure. Overall, the graph reflects the diverse landscape of imaging modalities used in BM assessment. The percentage distribution of utilization underscores the importance of considering factors such as sensitivity, specificity, availability, cost, and clinical context when selecting the most appropriate imaging modality for each case.

\section{Datasets}
\label{sec5}
In this section, datasets used in BM analysis are discussed. We begin the discussion with an overview of the characteristic features that distinguish the datasets from each other. These include the medical center where the data were collected, the imaging method used for acquisition, the number of patients, scanners, and images, the availability of the datasets, and potential biases related to the participants or their demographics. We then provide an overview of publicly available datasets of BM images that have been used in the literature to evaluate ML methods in the context of BM. In addition, we present proprietary datasets that can be used for future evaluation of BM identification methods. These datasets play a crucial role in the early identification and characterization of BM by extracting significant features. With these extracted features, various CADs can be performed to improve the accuracy and efficiency of diagnosis of bone metastases. This in turn leads to rapid interventions and contributes to better patient outcomes. To help researchers utilize these datasets effectively, we provide brief source acquisition information for each dataset, including the methodology used for data acquisition. This information is a valuable resource for researchers who want to understand the intricacies of data collection or create their own datasets.
Table \ref{tabdatasets} provides a summary of these datasets, including information such as the reference of the dataset, the number of subjects, the number of images, the modality used to detect bone lesions, the availability of the data, and other specific information.

\begin{table}
\small
\centering
\caption{The Summary and Characteristics of Reviewed Databases in BM analysis, BM = Bone Metastasis, ‘N/A’ in the table cells indicate not available information, BS= Bone Scintigraphy, CT= Computed Tomography,  }
\label{tabdatasets}
\begin{tabular}{p{2cm}p{1.5cm}p{1.5cm}p{1.8cm}p{1.8cm}p{4cm}}
\hline
\textbf{Dataset} & \textbf{\# of subjects} & \textbf{\# of images} & \textbf{Modality} & \textbf{Availability} & \textbf{Dataset Information} \\
\hline
\multicolumn{6}{c}{\textbf{Public Datasets}} \\ \hline
BS-80K \cite{huang2022bs} & 3247 & 82544 & BS & Yes & Images of the whole body and specific regions. \\
& & & & & Binary labels for classification. \\
& & & & & Bounding boxes for object detection. \\
\hline
BM-Seg \cite{afnouch2023bm} & 23 & 1517 & CT & Yes & Images of whole CT scan. \\
& & & & & Bone and lesion masks for segmentation. \\
\hline  
\multicolumn{6}{c}{\textbf{Private Datasets}} \\ \hline
Panpandrios et al. \cite{Papandrianos_2020a} & 507 & 586 & BS & Restricted access & Whole body images only.\newline{Binary and ternary labels for classification.}  \\
\hline
Pi et al. \cite{Pi_2020} & 13,811 & 15,474 & BS & Partly available with restricted access & Whole body images only.\newline{Binary labels for classification. }\\
\hline
Apiparakoon et al. \cite{apiparakoon2020malignet} & 9,280 & 19,648 & BS & No & Chest images only for lesion instance segmentation. \\
& & & & & The dataset contained 19,648 images of which 1,088 were labeled and the remaining 18,560 were unlabeled. \\
\hline
Han et al. \cite{Han_2021} & 5342 & 9133 & BS & No & A total of 2991 positive and 6142 negative from only prostate cancer patients. \\\hline
Shimizu et al. \cite{shimizu2020automated} & N/A & 246 & BS & No & Skeleton segmentation. \\\hline
Cheng et al. \cite{cheng2021lesion} & 576 & 576 & BS & No & Thoracic images only from 205 prostate cancer patients and 371 breast cancer patients. \\
& & & & & Binary labels for classification and bounding boxes for object detection. \\\hline
Masoudi et al. \cite{Masoudi_2021} & 114 & 2,880 & CT & No & Bone lesion classification purpose.
\newline{CT scans of 114 prostate cancer patients with 41 are metastatic.} \\\hline
Noguchi et al. \cite{noguchi2022deep} & 169 & 269 & CT & No & A total of 269 positive scans were collected including 1,137 BM and 4,63 negative scans for segmentation of the BM. \\\hline
Moreau et al. \cite{moreau2020deep} & 24 patients & N/A & PET/CT & No & Bone lesion segmentation and detection. \\\hline
Cao et al. \cite{Cao_2023} & 130 & 260 & SPECT & No & Thoracic images only. \\
& & & & & Bounding boxes for detection. \\\hline
Lin et al. \cite{Lin_2020} & 76 patients & 112 & SPECT & No & Thoracic images only of 76 patients diagnosed with metastasis. \\
& & & & & Lesion masks for segmentation. \\
\hline
\end{tabular}
\end{table}
\subsection{Public Datasets}
\textbf{ BS-80K \cite{huang2022bs}} emerged as a significant breakthrough in the field of datasets, addressing the lack of publicly available resources until November 2022. This dataset consists of a vast collection of 82,544 bone scan images obtained from 3,247 patients at West China Hospital. Each patient contributes two whole bone scan images, representing the anterior view (ANT) and posterior view (POST), along with 13 region-wise slices for susceptible body parts. Expert specialists meticulously annotated the images using an authorized labeling criterion, providing accurate labels for bone metastasis presence. Furthermore, multiple bounding boxes containing suspectable hot spots and their corresponding annotations are included within each body image. The availability of BS-80K has greatly facilitated research in bone metastasis, offering a comprehensive dataset that aids in algorithm development and analysis for improved detection and understanding of this medical condition.

\textbf{ BM-Seg \cite{afnouch2023bm}} consists of 23 CT scans of 23 patients with BM, including 9 female and 14 male subjects aged 18
to 83 years. All CT scans were retrospectively collected at UHC (University Hospital Centre) Hedi Chaker, Sfax, Tunisia. Data were collected from November 2020 to June 2022. Each CT scan was reviewed
by three experienced radiologists. Based on the location of bone pain, the patient’s health status, and any history of trauma, two physicians classified the slices of each CT scan as infected or not infected. Then, a radiologist (with more than 20 years of diagnostic imaging experience) reviewed the annotated slices that had originally been rated by the other two physicians. Finally, the radiologists selected the regions corresponding to the BM lesions in each infected slice of the CT scan. The examination was excluded if there was no clear agreement in diagnosis between the three radiologists. The slices that met the approval of all three radiologists were converted to JPEG format, then the bone and BM regions were manually segmented to extract ground truth masks (GT) using Apeer software. Since the labeling process is time-consuming, 70 infected slices were randomly selected from each CT-scan. On the other hand, for the CT-scans that had less than 70 infected slices, all infected slices were selected. A total of 1517 slices were annotated by creating the bone metastasis and bone
masks.

\subsection{Private Datasets}
Papapandrios and his team have conducted a series of studies aimed at proposing different datasets related to bone metastasis. In their first work \cite{Papandrianos_2020a}, they utilized a dataset consisting of 586 consecutive whole-body scintigraphy images of men, sourced from 507 different prostate cancer patients. These images were carefully selected and diagnosed by a nuclear medicine specialist. Out of the 586 bone scan images, 368 were from male patients with bone metastasis, while 218 were from male patients without bone metastasis. The nuclear medicine physician categorized all the patient cases into two categories: 1) metastasis absent and 2) metastasis present. Building upon their initial work, in their second study \cite{Papandrianos_2020c}, Papapandrios and his team aimed to expand the dataset by including additional bone scan images. They retrospectively reviewed a total of 778 planar bone scan images from patients with known prostate cancer. Of these, 328 bone scans were from patients with bone metastasis, 271 were benign cases with degenerative changes, and 179 were from normal patients without bone metastasis. The nuclear medicine physician classified these cases into three categories:  healthy, degenerative, and malignant. In another endeavor to gather more data on bone metastasis, Papapandrios, and his team \cite{Papandrianos_2020d} collected information specifically related to breast cancer as the primary tumor. They conducted a retrospective review of 422 consecutive whole-body scintigraphy images obtained from 382 different breast cancer patients (women). Out of the total 408 bone scan images the nuclear medicine specialist selected, 221 were categorized as malignant and 187 as benign, without bone metastasis. In \cite{Pi_2020}, Pi et al. utilized a vast dataset comprising more than 15,000 bone scan images from 13,811 patients to investigate the identification of bone metastasis. The dataset included 9595 benign diagnoses and 5879 malignant cases. The patients in the dataset consisted of 6699 males and 7112 females. However, currently, only the validation subset with limited access is available.

The dataset used in \cite{apiparakoon2020malignet} consisted of medical data from a total of 9,824 patients. The dataset was utilized for two main purposes: chest detection and lesion instance segmentation. For chest detection, 680 whole-body images were used for training, with 200 for validation and 240 for testing. In the case of lesion instance segmentation, the dataset consisted of 19,648 chest images, of which 1,088 were labeled and the remaining 18,560 were unlabeled.  Cheng et al. \cite{cheng2021lesion} conducted a study on BM identification using 524 whole-body bone scans collected from China Medical University Hospital. The dataset included 194 scans from prostate cancer patients and 371 scans from breast cancer patients. However, the study suffered from insufficient data as only 524 bone scan images were utilized.

The datasets provided by Shimizu et al. \cite{shimizu2020automated} and  Cheng et al. \cite{ cheng2021lesion} were characterized by their relatively small size with 246 and 524 bone scans,  respectively. In addition to bone scans, the previous works in the field incorporated various imaging modalities including CT, SPECT, and PET/CT. Using CT scans, Masoudi et al.\cite{Masoudi_2021} conducted a study in which they utilized 2,880 annotated bone lesions from CT scans of 114 patients diagnosed with prostate cancer. Among these patients, 41 had histopathologically confirmed metastatic bone lesions. Additionally, another study \cite{noguchi2022deep} collected a total of 269 positive CT scans, consisting of 1,375 bone metastases, and 463 negative scans for lesion segmentation purposes. Despite the limited number of patients and scanners available, the authors in \cite{Lin_2020, Cao_2023} used SPECT, a medical imaging technique, to detect bone metastases. While in \cite{moreau2020deep}, authors used  PET/CT images from 24 breast cancer patients to detect and segment automatically bones and metastatic bones.
\section{Machine Learning Tasks in Bone Metastasis}
\label{sec6}
\begin{figure}[h]
\centering
\includegraphics[width=0.9\textwidth]{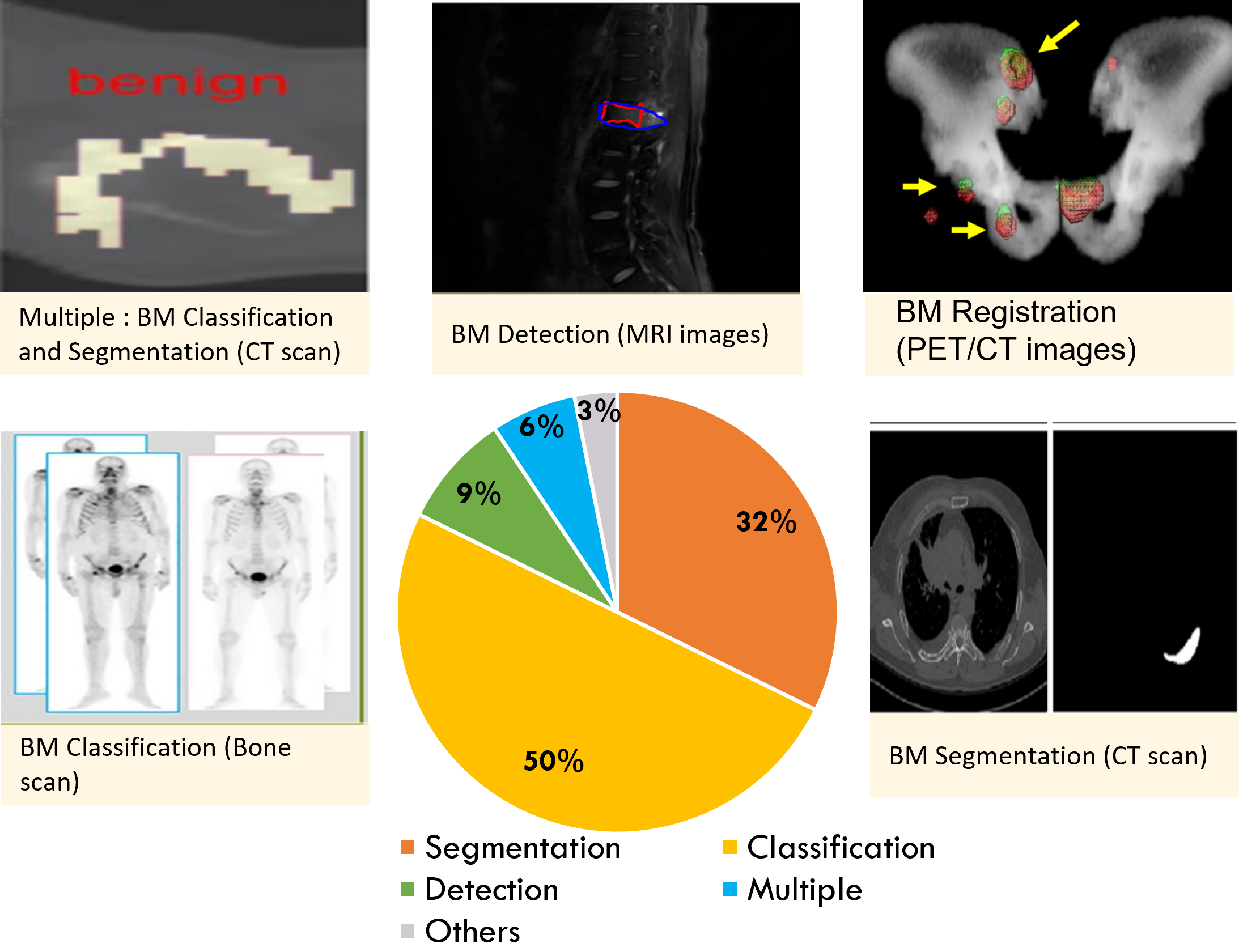}
\caption{A pie chart of the distribution of papers and visual examples of AI functionality. Examples are taken from papers as the 
following: Multiple\cite{Yildiz_Potter_2023}, detection \cite{wang2017multi}, registration \cite{yip2014use}, segmentation \cite{afnouch2023bm} and classification \cite{Papandrianos_2020a}}
\label{tasks}
\end{figure}

The advances that have been made in AI over the last decade have led to a dramatic increase in the accuracy of most computer vision applications \cite{patricio2023explainable}.Among these applications, BM analysis stands out as one that has achieved human-level accuracy in classifying various medical data \cite{meng2023artificial}. Furthermore, the categorization of research articles on BM analysis based on visual recognition tasks using ML techniques, as shown in Figure \ref{tasks}, highlights the significant impact and the widespread application of AI in various aspects of medical data analysis. In this section, the research papers are classified based on the addressed ML tasks including classification, segmentation, detection, multiple, and others. In addition to the main tasks, the remaining tasks are categorized as 'Others', which includes different tasks such as prediction and registration. While the multiple category refers to the paper that combines two or more tasks. Figure \ref{tasks} shows that most of the articles analyzed apply ML algorithms for the classification task, which accounts for 50\% of the articles. The objective here is to provide a comprehensive understanding of the research landscape in the field of BM analysis. In the following subsections, we detail existing work in this field and highlight the methods and results related to each specific ML task.

\subsection{ Machine Learning for BM Classification}

\begin{figure}[h]
\centering
\includegraphics[width=1\textwidth]{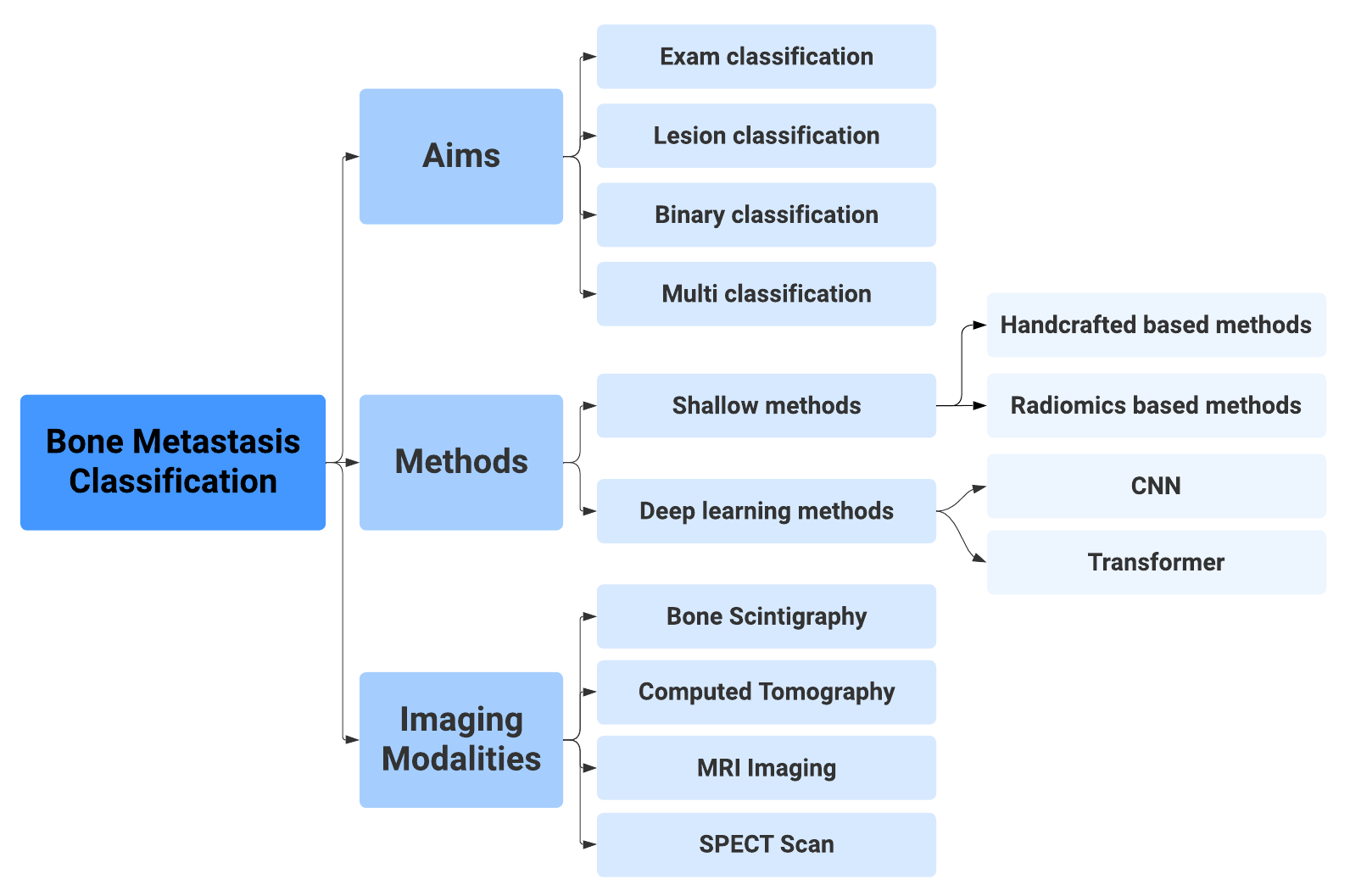}
\caption{Bone Metastasis Classification Overview }
\label{figbmclass}
\end{figure}

Image classification is a fundamental task in image recognition that aims to understand and categorize an image as a unified entity under a specific label \cite{li2023knowledge}. Classification of images or examinations is an area where ML plays an important role in BM analysis \cite{koike2023artificial}. BM classification can be briefly summarized in Figure \ref{figbmclass}.It includes different objectives, some studies focus on differentiating between healthy and metastatic patients or identifying the presence of metastases in an examination. Other studies, however, deal with the complicated task of classifying individual lesions such as osteolytic, osteoblastic or mixed. Furthermore, the diversity extends not only to the scope of the problem, but also to the used classification approach. While some researchers use binary classification schemes, others deal with the complexity of multiclass classification.

Various methods and techniques are used to achieve these goals. Originally, BM classification was performed using traditional image processing techniques such as manual feature detection and statistical classifiers\cite{Aslantas_2016,Koizumi_2020}. In addition, radiomics has emerged as an alternative method for BM classification, where the result is predicted by inputting manually defined texture and shape features extracted from a region of interest into ML models \cite{naseri2023scalable}. Furthermore, DL have been the dominant method in this field by using techniques such as convolutional neural networks (CNN) \cite{Papandrianos_2020a,belcher2017convolutional,deb2024cnn}. CNN can automatically compute the relevant features directly from the image data during training using a neural network architecture consisting of a series of convolutional and pooling operations.
In addition to the different classification approaches, five important imaging modalities are used in BM classification. These modalities, which include SPECT, CT, MRI and bone scintigraphy, offer different perspectives and insights into the pathology under investigation.
This section is organized by medical imaging modality and demonstrates the effective use of ML techniques for accurate and efficient BM classification. It includes a tabular summary of the proposed approaches and their performance metrics (Table \ref{tab_class}).

\subsubsection{BM Classification using Bone Scintigraphy}

As can be seen in Table \ref{tab_class}, bone scintigraphy is the most used medical imaging modality in classifying BM.
In fact, there are two main categories that use bone scintigraphy to classify BM, namely DL methods \cite{dang2016classification,
belcher2017convolutional, Ntakolia_2020, Papandrianos_2020a} and non-DL methods \cite{Koizumi_2017, Koizumi_2020,Aslantas_2016,Elfarra_2019,Calin_2021}. For non-DL methods, various computer-aided diagnosis (CAD) systems for BM classification have been proposed in numerous studies. BONENAVI® is one of these CAD systems, it stands out as a conventional approach that has shown reasonable sensitivity but variable specificity in detecting BM, which can prove useful in reducing false positives \citep{Horikoshi_2012, Tokuda_2014, Kikushima_2014, Koizumi_2014, Koizumi_2017, Koizumi_2020}. However, these non-DL methods' performance can be influenced by the primary cancer lesion.
Another automatic diagnostic system for identifying possible metastases in cancer patients was proposed by \citep{Aslantas_2016} baptized"CADBOSS". CADBOSS used image rasterization to extract features from whole-body
bone scans and then employed an ANN classifier at the lesion level.
On the other hand, the use of DL methods, especially CNN, has shown significant potential in the classification of BM. For example, Belcher \citep{belcher2017convolutional} has used CNNs to classify hot spots as benign or malignant. The hot spots on the lower spine were hand-extracted from prostate cancer patients and included in the study as they were considered the easiest to classify. In the same way, Dang et al. \citep{dang2016classification} used a patch-based CNN model to detect metastatic hot spots on bone scans. Similarly, some studies such as Papandrianos et al.\citep{Papandrianos_2020d, Papandrianos_2020c, Papandrianos_2020a, Papandrianos_2020b} and Zhao et al.\citep{Zhao_2020} with the basic architecture of CNN, which typically includes convolutional layers for feature extraction and pooling layers for downsampling, achieved higher accuracies of over 90\% and high AUC values for different cancer types.

\begin{figure}[h!]
\centering
\includegraphics[width=0.8\textwidth]{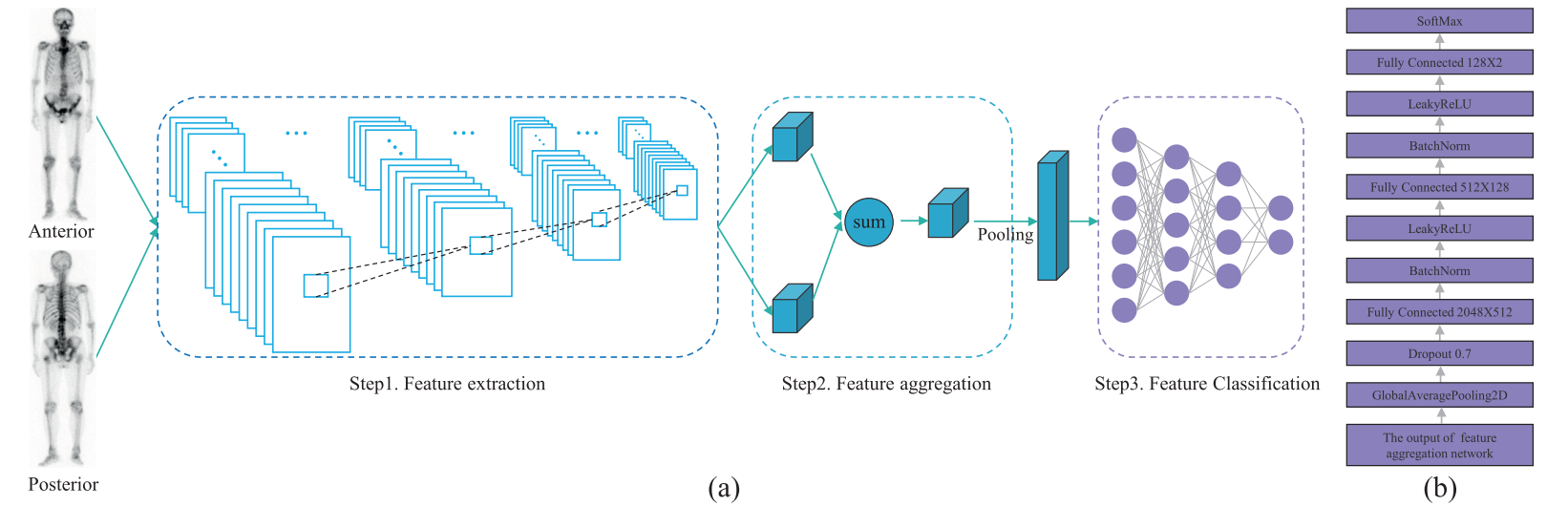}
\caption{Overall of the proposed three parts of the network proposed by Pi et al.\cite{Pi_2020}}
\label{pietalarchi}
\end{figure}

Due to the limited availability of data, numerous scientific publications have chosen transfer learning as a strategy to solve this problem. For example, Pi et al.\citep{Pi_2020} used the pre-trained model of Inception-V3 \cite{xia2017inception} to analyze anterior and posterior views in bone scans to detect BM in patients with different types of cancer. In this work, the authors employed a spatial attention feature aggregation operator to improve the extraction of spatial location information (see Figure \ref{pietalarchi}). 
In parallel, the authors in \citep{Papandrianos_2020e, Liuu_2021, Hajianfar_2021, Ibrahim_2023} attempted to use different pre-trained models to classify BM examinations.
These pre-trained CNN models have the potential to outperform conventional CAD systems and significantly reduce diagnosis time compared to human experts. However, the results showed that complicated models such as DenseNet did not perform better, as larger models are more likely to overfit due to the smaller amount of data.

When a large amount of training data is available, 2D-CNN modeling with a small number of trainable parameters is best suited for planar nuclear medicine scans such as bone scintigraphy \citep{Han_2021}. However, the use of DL is usually limited by the time-consuming effort required to create accurate labels for large datasets. To overcome this limitation, Han et al. \citep{Han_2021} proposed a 2D CNN classifier tandem architecture that integrates whole-body bone scans with local patches under the name of global–local unified emphasis (GLUE). The authors have demonstrated that creating local patches to increase the volume of sample data and integrating them with the sliding patches and conventional whole-body input can improve the performance of 2D CNN in a data-limited scenario while minimizing the risk of overfitting.

In addition, the increasing accessibility of large language models such as ChatGPT \cite{thirunavukarasu2023large}, coupled with their integration into medical image analysis tools, is paving the way for innovative advances in areas such as BM analysis. DL models in medical image analysis have shown the potential to improve diagnostic accuracy, but their implementation requires a combination of medical and programming skills \cite{moassefi2023reproducibility}. For this reason, Son et al. \cite{son2023chatgpt} conducted a study to investigate the feasibility of constructing a DL model using ChatGPT for BM diagnosis in bone scans and evaluate its diagnostic performance. They used ResNet50 as the backbone network and employed a class activation map algorithm for visualization to eliminate the imbalance between classes. However, it is worth noting that their proposed model primarily focused on scattered BM within the spine, which could potentially lead to confusion with benign spinal lesions or intense urinary activity.

\begin{longtable}[htbp]{p{0.35in}|p{1.59in}|p{1.59in}|p{2in}}
  \caption{Summary of existing Bone Metastasis classification works "AUC= Area under the ROC Curve, Acc=Accuracy, Sens=Sensitivity, Spec=Specificity, Prec=Precision, F1=F1-score, MM=Multiple Myeloma, conv= convolution layer, FC= Fully Connected, LR= Logistic Regression"}
  \label{tab_class} \\
  \hline
  \textbf{Ref.} & \textbf{Method} & \textbf{Dataset} & \textbf{Evaluation Metrics} \\ \hline
  \multicolumn{4}{c}{\textbf{Bone scintigraphy}} \\ \hline
    \cite{Koizumi_2020} & BONENAVI (ANN) & 54 patients & Sens=76\% \\
    \hline
     \cite{Aslantas_2016} & CADBOSS (ANN) & 60 patients,130 images & Spec=87\%,Sens=94\%,Acc=92\% \\
    \hline
    \cite{dang2016classification} & CNN & 2164 patients & Acc=89\%  \\
    \hline   
    \cite{belcher2017convolutional} & CNN & 2146 patients,10,428 hotspots & AUC=97.39\% \\
    \hline
    \cite{Elfarra_2019} & Parallelepiped Classification (PC) & 12 patients & Acc=87.58\%,Kcoefficient=0.8367 \\
    \hline
    \cite{Ntakolia_2020} & LB-FCN & 778 images & Acc=97.41\%  \\
    \hline
\cite{Papandrianos_2020a}  & CNN (3 conv layers, 1 dense layer) & 586 images & Acc=97.38\%,Sens=95.8\%  \\ \hline
\cite{Papandrianos_2020e}  & VGG16, DensNet & 778 scans & DensNet Acc=92.08\% $\pm$2.81\%,VGG16 Acc=92.14\% $\pm$ 2.91\% \\ \hline
\cite{Zhao_2020}  & CNN & 12,222 patients & AUC=95.5\%  \\ \hline
\cite{Pi_2020}& pre-trained Inception-V3 & 15,474 images,13,811 patients & F1=0.933,Acc=95.00\%, Sens=93.17\%,Spec=96.60\% \\ \hline
\cite{Liuu_2021}  & CNN (ResNet50) & 3352 patients & Acc=81.23\%, Sens=81.30\%, Spec=81.14\% \\ \hline
\cite{Calin_2021}& SVM and KNN & 9 images & KNN Acc=86.62\%, SVM Acc=86.81\% \\ \hline
\cite{Hajianfar_2021}& CNN pre-trained models & 2253 patients & AUC=68\%, Sens=66\%,Spec=71\%\\ \hline
\cite{Han_2021}  & 2 CNNs: WB model, GLUE model & 9113 bone scans,5342 patients & Acc=88.9\%(WB model), Acc=90\%(GLUE model) \\ \hline
\cite{Li_2022} & CNN & 2185 patients & F1=72.92\%,Acc=73.92\%, Prec=75.92\%,Sens=72.42\% \\ \hline
\cite{Ibrahim_2023} & VGG16 & 2365 images & Spec=80\%, Sens=82\% \\ \hline
\cite{son2023chatgpt}&ChatGPT 3.5 and  ResNet50& 400 BM patients,4226 normal patients& AUC=81.56\%, Sens=56\%, Spec=88.7\%\\
\hline
   \multicolumn{4}{c}{\textbf{SPECT}}
   \\ \hline
\cite{Lin_dspic_2021} & dSCIP (CNN) & 768 images, 384 patients & Acc=77.47\%,Prec=78.83\%, Sens=78.63\%,Spec=88.20\%, F1=78.60\%  \\
\hline
\cite{Lin_2021} & Dscint (CNN) & 600 patients, 1078 images & Acc=98.01\%, Prec=97.95\%,Sens=97.91\%, Spec=99.33\%,F1=97.92\%  \\
\hline
\cite{Linn_2021} & VGG ResNet, DenseNet & 251 Thoracic SPECT images & AUC=0.98 \\
\hline
\cite{Lin_2023} & 3 classifiers based on VGG16 & 3831 patients & Prec=92.9\%,Sens=96.6\%, F1=90.8\%,AUC=87.5\% \\
\hline
\cite{Chen_2021} & VGGNET & 642 images & Acc=99.20\%, Prec=99\%, Sens=99\%, F1=95\% \\
\hline
\cite{Zhao_2021} & VGGNET & 642 images & Acc=91.4\%, Prec=88.4\%, Sens=95.3\%, F1=99.5\%  \\
\hline
\cite{Liuff_2022} & CNN & 615 images & Acc=88.14\%, Prec=78.06\%, Sens=79\%, F1=78.99\% \\
\hline
\cite{Cao_2022} & Att-ResNet24 & 1668 images & Acc=73.7\%, Prec=74.4, Sens=73.6, F1=73.5  \\
\hline
\cite{wang2024automated} & self-defined CNN & 527 patients, 1054 images & Acc=80.38\%, Prec=80.51\%, Sens=80.39\%, Spec=80.39\%, F1=80.36\%, Auc=84.89\%  \\
\hline
   \multicolumn{4}{c}{\textbf{CT}}
   \\ \hline
\cite{Wiese_2012}& Watershed algorithm,Graph cut& 22 clinical cases  & Sens=71.2\%  \\
\hline
\cite{Roth_2015} & CNN & 59 patients& Sens= 60-80\%,  AUC=83.4\% \\
\hline
\cite{Mutlu_2021} & ML classifiers, best algo: KNN  &58 patients &Sens=82.4\% Spec=81.5\%,Acc=82.4\%, AUC=0.861 \\
\hline
\cite{Masoudi_2021} & CNN (2D ResNet50,3D ResNet182D)  & 2.880 CT scans,114 patients & Acc= 92.2\%, F1=92.3\%\\
\hline
\cite{Naseri_2022} &Radiomics+ML classifiers &359 patients & AUC=0.97 for GPR (best classifier) \\
\hline
\cite{naseri2023scalable} & Radiomics +NLP,ML classifiers &176 patients  &Acc=0.82, Sens=0.59, Spec=0.85, AUC=0.83\\
\hline
\cite{koike2023artificial} &  YOLOv5m, InceptionV3    &2125 images, 79 patients& Acc=0.872, prec=0.948, Sens=0.741, F1=0.832, AUC=0.941  \\
\hline
\cite{Acar_2019} & ML Classifiers  &75 PET/CT patients&  AUC=0.76 for best classifier: weighted KNN algorithm \\
\hline
\cite{afnouch2023automatic} & Pretrained Transformer( ViT Tiny) and pretrained CNN (ResNext50) & \text{BM-Seg dataset} & Acc=86.94\%, F1=86.57\% \\
\hline
\multicolumn{4}{c}{\textbf{MRI}}
 \\ \hline
\cite{Xiong_2021} &  ML classifiers & 107 patients,60 MM lesions, 118 metastatic lesions& Acc=81.50\%,Sens=87.90\%, Spec=79\% \\
\hline
\cite{Chianca_2021} & Radiomics, ML classifiers & 146 patients& 2-label classification:Acc=86\%, 3-label classification:Acc=69\% \\
\hline
\cite{Liuuu_2021} & LR & 103 MM patients, 138patients with metastases & AUC=85\% \\
\hline
\cite{Hallinan_2022} & DL based on ResNet 50   &164 patients, 215 thoracic MRI spine& Kappas=0.94–0.95,p<0.001\\
\hline
\end{longtable}

\subsubsection{ BM Classification using SPECT}
Among the various medical imaging modalities used in BM classification, SPECT imaging is a prominent and widely used as indicated in Table \ref{tab_class}. Using this imaging modality, Lin et al. \cite{Lin_dspic_2021} have proposed different methods for BM classification. They presented a deep network called dSPIC for automatic multi-disease and multi-lesion diagnosis. The network was designed to extract optimal features from images and classify them such as metastasis, arthritis, and normal. Then, they proposed a self-defined CNN called Dscint \cite{Lin_2021}, which uses hybrid attention mechanism to classify whole-body SPECT images into different disease categories. They included AlexNet, ResNet, DenseNet, VGGNet, and Inception-v4 as backbones to make multiclass classifications. The accuracy of their proposed Dscint outperformed several famous deep classification networks with a value of 98.01\%. Based on popular deep networks which are ResNet, VGG, and DenseNet, they also developed deep classifiers \cite{Linn_2021} to automatically diagnose metastases in 251 thoracic SPECT bone images. More recently, they proposed three distinct two-class classifiers in \cite{Lin_2023} based on the VGG 16 model. These classifiers autonomously identify whether or not a SPECT image contains lesions.

In another study, Zhao et al.\cite{Zhao_2021} proposed a CNN-based classification model for accurate diagnosis of bone metastases. They adopted the standard VGG model to develop a classifier for SPECT images. In \cite{Chen_2021}, an automated classification model based on the VGG model was proposed to determine whether an image contains lesions or not. Liu et al. \cite{Liuff_2022} presented a method for classifying multiple diseases including normal cases, bone metastases, arthritis, and thyroid cancer using a customized CNN model called SPNT9. More recently, \cite{Cao_2022} focused on thorax image classification using a self-defined CNN called Att-ResNet24 with a hybrid attention mechanism. 
Recently, the authors in  \cite{wang2024automated} proposed an innovative approach to automatically diagnose BM using SPECT images through deep learning-based image classification. Their method involves a CNN architecture with distinct feature extraction and classification sub-networks tailored for detecting lung cancer BM. By analyzing SPECT bone scintigrams, the model achieved high detection accuracy, particularly when combining anterior and posterior scans while excluding the bladder.
 
\subsubsection{ BM Classification using CT scan}

Similar to bone scans, a well-established nuclear medicine imaging technique for the classification of bone lesions, CT scans are also used for this purpose. The classification method used with this technique can be either radiomics-based methods or non-radiomics-based methods. Among the studies included in this review that used non-radiomic methods, the work in \cite{Wiese_2012} presented a CAD system for the detection of sclerotic BM in the spine. The system used a watershed algorithm and graph cut to detect lesions and an Support Vector Machine (SVM) classifier \cite{cortes1995support} to classify them. Similarly, Roth et al. \cite{Roth_2015} proposed a two-stage coarse-to-fine cascade \cite{lu2011coarse} to detect sclerotic spinal metastases. Using CNN classifiers, difficult false positives were eliminated by an additional selection process. Later, Mutlu et al. \cite{Mutlu_2021} used eight ML algorithms to classify osteolytic bone lesions from 58 patients as malignant or benign. They found that KNN had the best predictive performance. Additionally, a DL model \cite{Masoudi_2021} using 2,880 annotated CT scans of 114 patients with prostate cancer was able to detect bone lesions and classify them as malignant or benign with an accuracy of 92 2\%.

A radiomic model that distinguishes BM from normal bone marrow was published in \cite{Naseri_2022}. After the lesions are manually located, they are automatically segmented to collect enough information. In the study, BM was successfully distinguished from healthy bone regions using various radiomic features and ML classifiers. More recently,
Naseri et al. \cite{naseri2023scalable} presented an approach combining NLP and radiomics to discriminate between painful and painless BM lesions in simulated CT images of cancer patients. By automatically extracting pain scores from clinical notes using NLP and identifying BM central points on CT images, the study extracted radiomics features from these areas. Koike et al. \cite{koike2023artificial} presents a DL-based CAD system for the detection and classification of lytic spinal metastases using CT scans. The system utilizes YOLOv5m (You Only Look Once) \cite{redmon2016you} architecture for vertebra detection and Inception-V3 architecture with transfer learning for lytic lesion classification.  DL models demonstrated high precision, precision, recall, and F1 score to detect and classify the presence of lytic lesions, with an average Intersection over Union (IoU) value of 0.923 for vertebra detection. The Grad-CAM technique was employed for visual interpretation, producing heat maps consistent with the location of lytic lesions.
Using PET /CT images, Acar et al. \cite{Acar_2019} focused on the use of CT texture analysis in combination with ML methods to distinguish between metastatic bone lesions and fully addressed sclerotic areas in prostate cancer patients with BM. With an AUC of 0.76, they found that the weighted KNN acheived the best performance.

\subsubsection{ BM Classification using MRI}

MRI is not commonly utilized for classifying metastatic bone like bone scan and CT scan due to its higher cost, as stated in a study by Chianca et al. \cite{Chianca_2021}. They analyzed the diagnostic performance of machine learning (ML) classifiers in differentiating spinal lesions using radiomic data. The lesions were categorized into two groups: benign and malignant (2-label classification) or three groups: benign, primary malignant, and metastatic (3-label classification). In a similar vein, Liu et al. \cite{Liuuu_2021} and Xiong X et al. \cite{Xiong_2021} used conventional T1-weighted (T1W) and fat-suppressed T2-weighted (T2W) magnetic resonance sequences to distinguish between spinal metastases and multiple myeloma (MM). Xiong et al. \cite{Xiong_2021} combined radiomics models with various ML algorithms to predict the likelihood of spinal metastases.  Liu et al. \cite{Liuuu_2021} used logistic regression to develop a model that performed well with 10-EPV (events per independent variable) in distinguishing MM from spinal metastases, with an area under the curve (AUC) of 0.85. Most recently, Hallinan et al. \cite{Hallinan_2022} designed an automated classification system for metastatic epidural spinal cord compression (MESCC) using the Bilsky classification on MRI.

\subsection{Machine Learning for BM Segmentation}

\begin{figure}[htbp]
\centering
\includegraphics[width=1\textwidth]{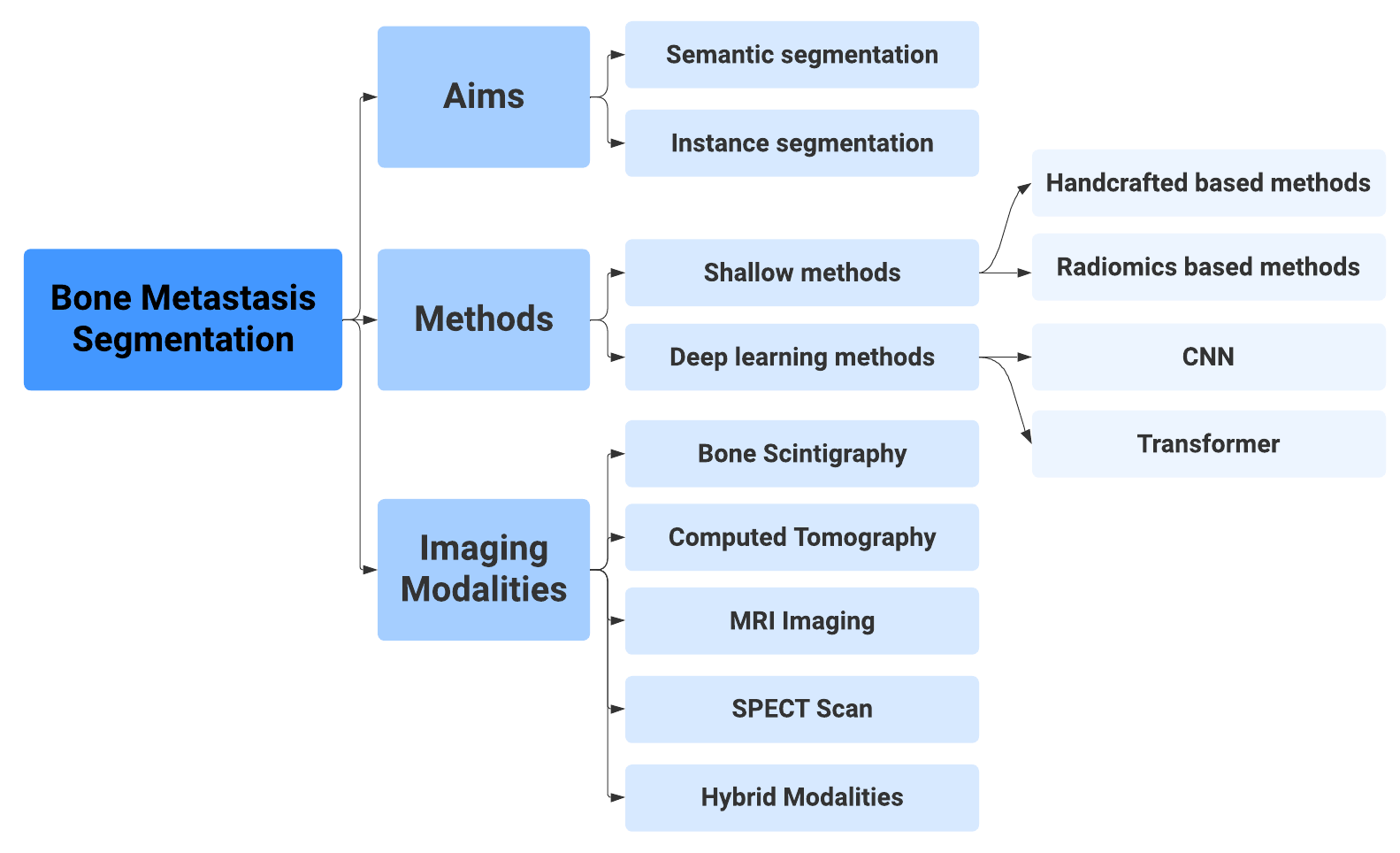}
\caption{Bone Metastasis Segmentation Overview }
\label{figbmseg}
\end{figure}

In recent years, the research on BM segmentation has increased significantly \cite{paranavithana2023systematic}. This increasing use reflects the growing realization that they are essential for accurate diagnosis and treatment planning \cite{belal2019deep}. BM segmentation involves the delineation and isolation of BM on medical images to allow a more accurate and quantitative assessment of these lesions \cite{paranavithana2023systematic,arends2022clinical}. Figure \ref{figbmseg} shows an overview of BM segmentation. Research efforts in BM segmentation can be broadly divided into two categories: semantic segmentation and instance segmentation. Semantic segmentation focuses on the identification and delineation of the entire region with BM in an image \cite{Gao_2020,Cao_2023}, while instance segmentation aims to differentiate and localize individual BM lesions \cite{apiparakoon2020malignet}. In addition, a variety of imaging modalities have been explored, including bone scans, CT scans, magnetic resonance imaging, and hybrid imaging modalities, especially PET / CT scans, for BM segmentation using different ML approaches. These approaches encompass DL methods \cite{Wu_2023,shimizu2020automated,Wu_2023,Shimada_2023,Saito_2021} and shallow methods \cite{Aslanta__2014,Gao_2020}.  

This section is dedicated to offer a comprehensive overview of the current state-of-the-art in Bone Metastasis segmentation. It is organized into subsections, each focusing on a specific imaging modality and providing detailed insights into the methods and algorithms developed for that particular modality. The section is concluded with a comprehensive tabular summary (Table \ref{tabseg}), highlighting the proposed approaches and their corresponding performance metrics.

\subsubsection{BM Segmentation using Bone Scintigraphy}

Consistent with the trends observed in BM classification, bone scans emerge as the most commonly used imaging modality for BM segmentation, as shown in Table \ref{tabseg}. While DL approaches predominate in this area, and UNet and its variants are widely used \cite{shimizu2020automated,Wu_2023,Shimada_2023,Saito_2021}, other methods are also becoming more prevalent. These include shallow techniques such as unsupervised approaches, such as the fuzzy C-Means clustering method \citep{Aslanta__2014}, which used to determine the locations and areas of bone involvement and anomalies. 

Using DL techniques, Shimizu et al. \citep{shimizu2020automated} proposed an image interpretation system for skeletal segmentation and extraction of BM hotspots. Similarly, Wu et al. \cite{Wu_2023} proposed a DL method that uses a Swin transformer \cite{liu2021swin} as a decoder to extract feature information from an image. Although the authors found that the Swin transformer has demonstrated an excellent ability to capture remote information in various domains, it could further improved, as shown by the IoU results. Recently, Shimada et al. \citep{Shimada_2023} presented a fine-tuning simulation of a post-market computerized bone scintigram detection system and analyzed its performance. They identified the factors that affect performance changes and provided useful information to develop an effective design scheme for continuous learning in AI systems. The double U-net model was adapted in \cite{Chen_2023} by incorporating background removal, addition of negative samples, and transfer learning methods for BM segmentation. These DL-based works used the UNet model \cite{ronneberger2015u}  as the basic architecture for their proposed approaches. However, the differences in the results obtained can be primarily attributed to differences in data quality and the size of the dataset, which emphasizes the central role of data features in influencing the results.

\subsubsection{BM Segmentation using SPECT}

In addition to bone scans, SPECT images are frequently utilized for the segmentation of metastatic bones. Although a variety of methods can be employed for this purpose, including both deep learning (DL) and non-DL techniques, the use of DL-based approaches has increased substantially due to their superior ability to capture intricate spatial relationships and image features relevant to the delineation of BM \cite{long2015fully}.
In \cite{Lin_2020}, Lin et al. used two different DL architectures, U-Net and Mask R-CNN. The results indicate that the constructed segmentation models attain a value of 0.9920 for PA, 0.7721 for CPA, 0.6788 for Rec, and 0.6103 for IoU. Similarly, Zhang et al. \cite{zhang2021bone} used the U-Net algorithm for segmentation of BM in SPECT images. Enhancing the UNet architecture with an attention mechanism in the skip connections has demonstrated its effectiveness in refining the selection of relevant data features while preventing feature redundancy. This modification has rendered the model more suitable for training. Empirical results have revealed that the proposed approach achieved an Intersection over Union (IoU) of 0.633 and a Dice Similarity Coefficient (DSC) of 0.571.

Another promising work was done by Che et al. \cite{Che_2021}, 
where they proposed to incorporate Unet \cite{ronneberger2015u} with attention mechanism to precisely segment the BM in the pelvic region. The experimental results demonstrated the effectiveness of their approach, which resulted in considerable improvements in BM segmentation compared to AttUnet and Res-Unet models. Additionally, Gao et al. \cite{Gao_2022} introduced a new UNet variant with an enhanced design for the automatic BM segmentation in lung cancer patients. This model integrates a fused residual structure and an attention mechanism. The fused residual structure replaces each block in the UNet architecture with a modified structure that consists of two 3×3 convolutional layers, processes and captures image features, and  a 1×1 convolutional layer, provides a shortcut connection to preserve important information. The attention mechanism selectively focuses on relevant information, thereby improving the model's accuracy and performance in BM segmentation.
\begin{figure}[h!]
\centering
\includegraphics[width=0.8\textwidth]{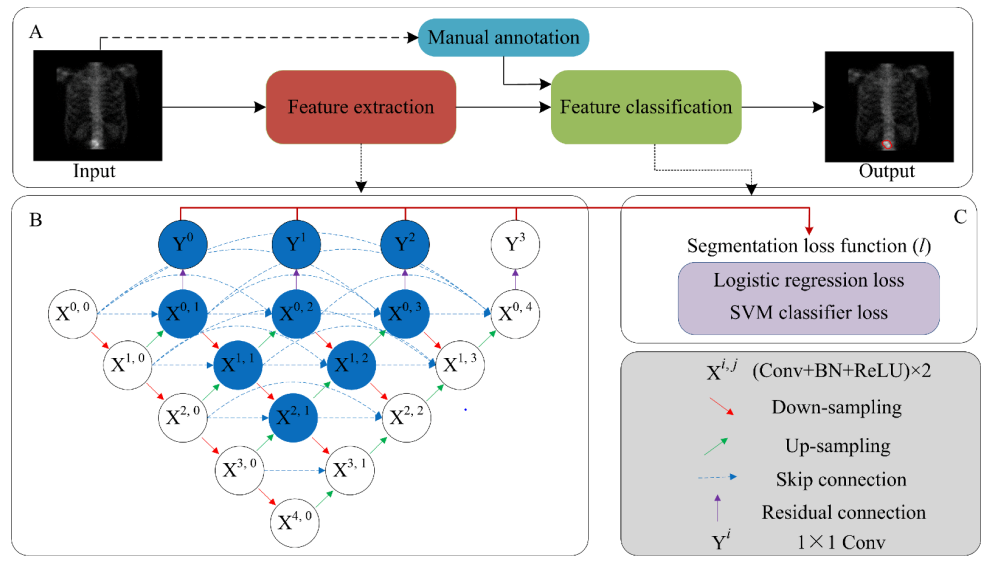}
\caption{(A) Framework proposed by Cao et al\cite{Cao_2023}, which consists mainly of feature extraction (B) and feature classification (C)}
\label{caoetalarchi}
\end{figure}
Recently, Cao et al. \cite{Cao_2023} presented a method that used a five-layer segmentation network based on the U-Net++ model, which was self-defined. Furthermore, they developed a view aggregation technique that utilized pixel-wise addition to enhance regions with high uptake of radiopharmaceutical while keeping the background area unchanged. The model comprised two sub-networks: feature extraction and pixel classification, and followed an encoder-decoder structure with deep supervision as shown in Figure \ref{caoetalarchi}. The feature extraction subnetwork extracted hierarchical features from the bone scan images, while the pixel classification sub-network identified and delineated pixels in the metastatic areas. Similarly, Xie et al.  \cite{xie2023segmentation} proposed an automated segmentation model based on the U-Net++ model, incorporating feature fusion and attention mechanisms to improve feature learning in crucial regions. The proposed model achieved a Dice similarity score of 0.6221 and a sensitivity score of 0.5878. Both studies have shown the effectiveness of DL techniques for the accurate segmentation of the lesion. Cao et al. \cite{Cao_2023} emphasized the importance of including more bone scan images to further improve the performance of the model, while Xie et al. \cite{xie2023segmentation} highlighted the benefits of integrating attention gates and feature fusion for better segmentation results.

\begin{figure}[h!]
\centering
\includegraphics[width=0.8\textwidth]{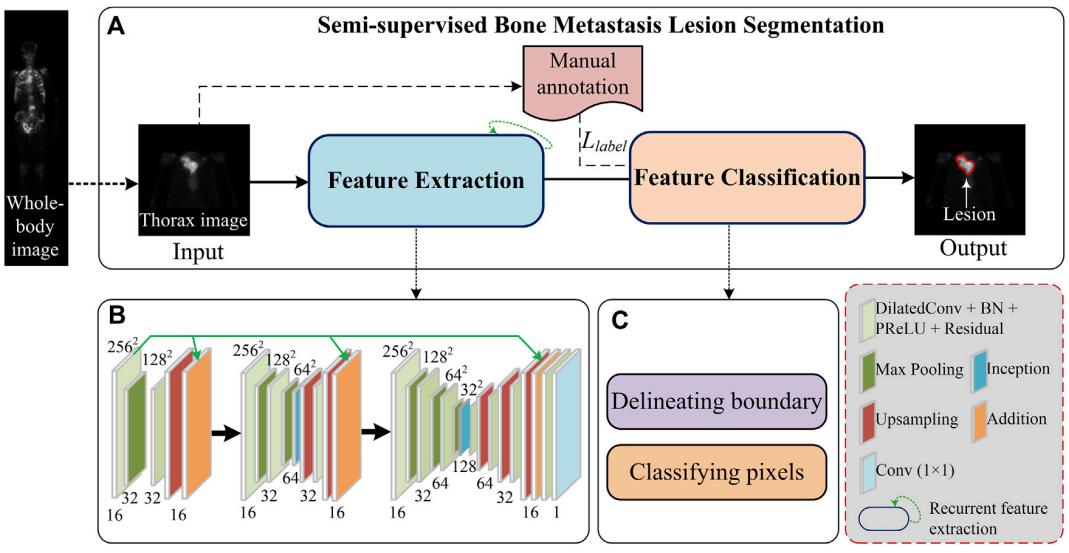}
\caption{Overall structure of the semi-supervised metastasis lesion segmentation system Lin et al\cite{Lin_2022}}
\label{lietalarchi}
\end{figure}

In contrast to previous work based on a supervised segmentation model, Lin et al. \cite{Lin_2022} proposed a user-defined semi-supervised segmentation model for identifying and delineating lesions of skeletal metastases. The proposed system consists of two modules as shown in Figure \ref{lietalarchi}. The first module extracts features and the second is for pixel classification. The feature extraction stage uses cascaded layers, including dilated residual convolution, inception connection, and feature aggregation, to learn hierarchical representations from low-resolution SPECT images. The pixel classification stage classifies each pixel into categories in a semi-supervised manner and delineates pixels belonging to an individual lesion using a closed curve. 
Similarly, Huang et al.\cite{Huang_2022} developed an end-to-end multi-task DL model for automatic lesion detection and anatomical localization in whole-body bone scintigraphy. The model architecture consists of two task flows: lesion segmentation and skeleton segmentation. The lesion segmentation flow is trained in a supervised manner using image patches, while the skeleton segmentation flow employs a semi-supervised approach. By jointly utilizing these flows through shared encoder layers, the model captures a more generalized distribution of the sample space and extracts abstract deep features.

In their study, Gao et al.\cite{Gao_2020} took a different approach to the predominant use of DL methods. Instead, they investigated three different algorithms, namely the K-Means clustering method, the region growth method, and the C-V model, to segment bone lesions. Interestingly, their comparative results showed that the C-V model outperformed the other two methods with 0.8076 for the Tanimoto similarity coefficient, suggesting that it has the potential to significantly assist oncologists.

\subsubsection{BM Segmentation using CT}

\begin{figure}[h!]
\centering
\begin{minipage}{0.8\textwidth}
\hspace{2cm}
\includegraphics[width=\textwidth]{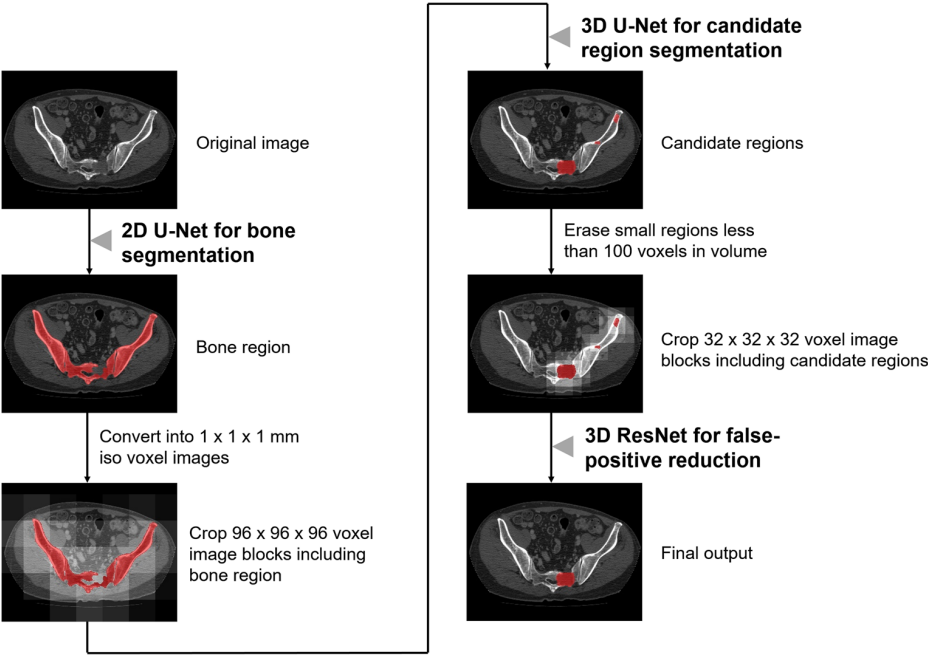}
\end{minipage}
\caption{Schematic of the proposed algorithm in Noguchi et al. \cite{noguchi2022deep}}
\label{noguchiarchi}
\end{figure}

BM segmentation in CT scans is primarily based on DL methods, in particular the UNet model \cite{ronneberger2015u}, which was used in the study by Chang et al. \cite{Chang_2021} for segmentation of BM in spinal sclerosis on CT images. The authors acknowledged the potential of CNNs in aiding lesion detection while recognizing the need for further refinement and broader validation. Subsequently, Noguchi et al. \cite{noguchi2022deep} developed a segmentation approach utilizing three CNNs: a 2D UNet-based network for bone segmentation, a 3D UNet-based network for candidate region segmentation, and a 3D ResNet-based network for reducing false positives as shown in Figure \ref{noguchiarchi}. The study demonstrated that the algorithm improved radiologists' overall performance in detecting BM while reducing interpretation time. Recently, Chang et al. \cite{Chang_2023} discussed the use of the UNet model for the automatic detection of lytic lesions of the spine in chest, abdominal and pelvic CT images. The authors manually segmented the images into three masks: lesion, normal bone, and background. The trained model yielded promising results with a mean Dice score of 0.61 for lytic lesions, 0.95 for normal bone, and 0.99 for background.
In parallel, Afnouch et al.\cite{afnouch2023bm} presented a novel CNN-based approach for BM segmentation that introduces the hybrid AttUnet++ architecture with dual decoders for simultaneous segmentation of BM and bone regions. Furthermore, they utilized an ensemble of trained hybrid AttUnet++ models (EH-AttUnet++) to improve segmentation performance and showed superior results compared to existing approaches on several evaluation metrics.

In contrast to the UNet-based methods mentioned in previous studies, Song et al.\cite{Song_2019} employed a modified Holistic Nested Edge Detection (HED) network. The authors emphasized the challenge that noisy CT images present in accurately classifying BM and detecting lesions in specific regions. To address this issue, they proposed a segmentation method that used the HED network to identify the contour and location of injury areas on CT images. Motahashi et al. \cite{motohashi2023new} have addressed the automatic segmentation of osteolytic BM lesions in the thoracolumbar region through the development of a new DL-based computer-aided detection model using DeepLabv3+ \cite{chen2018encoder}. DeepLabv3+ incorporates the Atrous Spatial Pyramid Pooling convolutional layer, enabling the extraction of a wider range of image features and facilitating accurate detection of objects, even in cases involving smaller areas or regions with indistinct boundaries. The utilization of DeepLabv3+ demonstrates promising progress in automating the identification of osteolytic BM lesions, potentially enhancing diagnostic efficiency and contributing to improved treatment strategies for individuals with thoracolumbar BM.
These studies demonstrate the prevalence of DL methods, particularly the UNet model, in BM segmentation on CT scans. Additionally, the use of methods such as DeepLabV3+ highlights the diverse techniques employed to improve the accuracy and efficiency of BM segmentation.


\subsubsection{ BM Segmentation using MRI}

\begin{figure}[h!]
\centering
\includegraphics[width=0.8\textwidth]{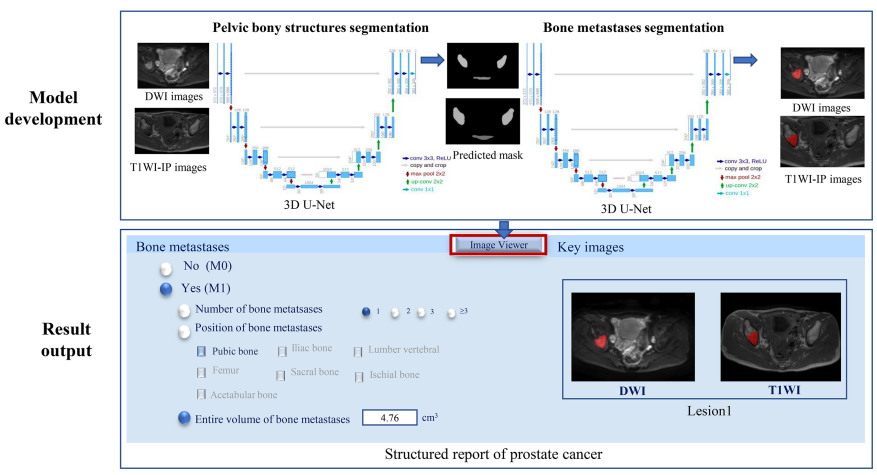}
\caption{A Two-Step 3D U-Net Approach for BM Segmentation proposed by Liu et al. \cite{ Liu_Xiang_2021}}
\label{liuarchi}
\end{figure}

In the field of medical imaging, MRI has proven to be a valuable tool for BM segmentation, despite its higher cost compared to alternative modalities such as bone scans and CT scans. MRI's superior sensitivity and specificity in detecting and characterizing metastatic lesions make it an attractive option for BM segmentation. Recent advancements in DL techniques and innovative architectures have significantly improved the accuracy and precision of BM segmentation on MRI.
For example, Liu et al. \cite{ Liu_Xiang_2021} proposed a two-step method for segmenting BM, with a particular focus on pelvic BM. This methodology utilized dual 3D UNet architectures (see Figure \ref{liuarchi}), incorporating T1-weighted and diffusion-weighted images, to accurately identify and segment metastatic lesions within the pelvic bony structures. The initial step of their approach involved using a 3D UNet algorithm for pelvic bone segmentation. The purpose of this stage is to accurately demarcate the structures of the pelvic bone, providing a precise anatomical context for the subsequent segmentation of the metastatic lesions. Following the successful segmentation of the pelvic bone, the researchers proceeded to the second step, which involved employing another 3D UNet model specifically designed for the segmentation of bone lesions within the previously segmented pelvic bony structures.
Subsequently, Hille et al. \cite{hille2020spinal} investigated the automatic segmentation of vertebral metastases in MRI with Deep CNN, in particular with a UNet architecture. Their results showed that automatic segmentation is almost as accurate as expert annotation with a DSC of up to 0.78 and a mean sensitivity of up to 78.9\%. This study underscores the potential of DL algorithms for accurate and reliable segmentation of vertebral metastases.
To further improve the accuracy of spinal metastases segmentation, Wang et al. \cite{wang2023contrast} focused on the use of contrast-enhanced magnetic resonance imaging (CE-MRI) in combination with advanced DL architectures. Their proposed approach combines a region-growing algorithm for lesion segmentation with an improved U-Net with Inception-ResNet model. The results showed improved diagnostic accuracy, with the Radiomics approach achieving a dice score of 0.74 and improving U-Net performance with achieving an average average diagnostic accuracy of 0.98. The precision accuracy of their approach (0.98) demonstrating its potential to help with the differential diagnosis of spinal metastases and primary malignant bone tumors of the spine.

In conclusion, despite the higher costs associated with MRI compared to other imaging modalities, its use in BM segmentation  has proven beneficial due to its superior sensitivity and specificity. Incorporation of DL techniques and innovative architectures has further improved the  precision of bone metastasis segmentation, as demonstrated by the studies mentioned above. These advances contribute to the growing body of knowledge in the field of medical imaging and promise to improve clinical decision-making and patient outcomes in the treatment of BM.

\subsubsection{BM Segmentation using hybrid imaging modalities}
In addition to the use of single medical imaging modalities, Hybrid approach has get more attention in the recent years for BM segmentation. Unlike BM classification, which mainly rely on using single imaging modalities, hybrid modalities have shown an impressive performance along side using DL. The most famous hybrid imaging modalities for BM segmentation is the Positron Emission Tomography/Computed Tomography (PET/CT) .   
In a study conducted by Moreau et al. \cite{moreau2020deep}, the effectiveness of two approaches for segmenting bone lesions in metastatic breast cancer is compared with nnUnet performance \cite{Isensee_2020}. Their first approach involves using lesion annotations of PET and CT images as a two-channel input, while the second method utilized both reference bone masks and lesion masks as ground truth. Inclusion of bone masks resulted in improved precision and a slight increase in the dice score for bone lesion segmentation.

In another study by Murugesan et al. \cite{Murugesan_2023}, the use of DL algorithms for automated PET/CT lesion segmentation in oncology was investigated. The authors highlighted the challenges associated with manual tumor annotation in whole-body PET/CT scans and emphasized the potential of DL-based automated tumor segmentation. Their approach achieved a dice score of 0.5541 on the test split, which comprised 150 cases. These studies jointly emphasize the importance of hybrid imaging methods, particularly PET/CT for BM segmentation. They highlight the potential of cutting-edge DL techniques for precise and automated segmentation of bone lesions, providing valuable insight for improving the diagnosis and treatment response.

\begin{longtable}[htbp]{p{0.35in}|p{1.59in}|p{1.59in}|p{2in}}
  \caption{Summary of existing Bone Metastasis segmentation works "DSC= Dice similarity coefficient, Acc= Accuracy, Sens=Sensitivity, Spec= Specificity, Prec=Precision F1= F1-score, N/A=not applicable, GFI=The goodness of fit index, Tanimoto= Tanimoto coefficient similarity,  MIoU= average cross-merge ratio, PA= Pixel Precision, FPR=False positive regions, CPA= Class Pixel Accuracy, Jaccard= Jaccard Similarity Coefficient"}
  \label{tabseg} \\
  \hline
  \textbf{Ref.} & \textbf{Method} & \textbf{Dataset} & \textbf{Evaluation Metrics} \\ \hline
  \multicolumn{4}{c}{\textbf{Bone scintigraphy}} \\ \hline
\cite{Aslanta__2014}& Fuzzy C-Means& 12 patients & N/A \\
\hline
\cite{shimizu2020automated} & BtrflyNet & 246 patients & DSC = 0.8756 \\
\hline
\cite{Saito_2021} & EnsembleNet: ResBtrflyNet+ ResBtrflyNet & 665 patients,330 scans& FPR= 2.13 for anterior,2.62 for posterior \\
\hline
\cite{Huang_2022} & Semi-supervised DL model& 617 patients  & DSC=89.20, Prec = 81.32,Sens=  90.67 \\
\hline
\cite{Wu_2023}& Swin Transformer& 242 images & Acc=97.65\%, DSC=77.81\% ,IoU= 35.59\% \\
\hline
\cite{Shimada_2023}  & ResBtrflyNet &1032  patients from 5 hospitals & GFI= 0.862 \\
\hline
\cite{Chen_2023}&   Double UNet & 100 breast cancer,100 prostate cancer &Prec = 63.08\%,Sens= 70.82\%, F1=66.72\% \\
\hline
\multicolumn{4}{c}{\textbf{SPECT}} \\ \hline
\cite{Gao_2020}&  K-means, region growth,C-V model & 120 images & Tanimoto= 0.7307,Tanimoto=0.7768,Tanimoto=0.8076\\ 

\hline
\cite{Lin_2020}& U-Net,Mask R-CNN& 112 images,2.280 augmented, 76 patients & Acc= 0.9920,CPA= 0.7721, Sens= 0.6788, IoU=0.6103 \\
\hline
\cite{zhang2021bone}& Attention U-Net   &125 images &  DSC=0.571 \\ 
\hline
\cite{Che_2021}& Attention U-Net & 121 images, 2390 sheets(data augmentation)& IoU=0.6045,DSC= 0.7214, Prec= 0.7564 \\
\hline
\cite{Gao_2022} & UNet  & 306 diagnostic records, 176 patients &
 MPA = 0.7824 ,PA = 0.9955 , MIoU = 0.7291 \\ 
\hline
\cite{Lin_2022}  & Semi-supervised model&724 whole-body images,362 patients & DSC= 0.692\\

\hline
\cite{Cao_2023}   &Model based on UNet++ &130 patients, 260 images  & DSC=0.6556,Sens= 0.6257, CPA= 0.6885\\
\hline
\cite{xie2023segmentation}   & Model based on UNet++ & 306 images&   DSC=0.6221 ,CPA = 0.6612, Sens= 0.5878 
\\
\hline
\multicolumn{4}{c}{\textbf{CT}} \\ \hline
\cite{Song_2019}& modified HED Network&  21 patients,250 images &  TP rate=79.8\% ,IOU= 69.2\%  \\
 \hline
\cite{Chang_2021} & Deep CNN & 600 images  & DSC=0.83 for lesion,0.96 for normal bone,0.99 for background\\
\hline
\cite{noguchi2022deep}& CNN, 3D UNET, 3D Resnet & 732 patients  &Sens = 82.7\%, FP= 0.617\\
\hline
\cite{Chang_2023}& UNET & 600 images & DSC = 0.61 for lytic lesion,0.95 for normal bone,0.99 for background \\
\hline
\cite{motohashi2023new}& DeepLabv3+ & 475 scans& Sens=0.78, Prec=0.68, F1=0.72 \\
\hline
\multicolumn{4}{c}{\textbf{MRI}} \\ \hline
\cite{Liu_Xiang_2021}& 3D UNet & 859 patients &DSC>0.85,Hausdorff
distance <15 mm \\
\hline
\cite{hille2020spinal} & UNet & 40 patients &DSC= 77.6\%, Sens= 78.9\%\\
\hline
\cite{wang2023contrast} & Improved UNet, Inception-ResNet  & 81 patients, 81 scans& PA=98.001\%,IoU=96.819\%,DSC =98.384\%\\
\hline
\multicolumn{4}{c}{\textbf{PET/CT}} \\ \hline
\cite{moreau2020deep}& nnUnet & 24 patients& DSC= 0.61 \\
\hline
\cite{Murugesan_2023} & residual 3D Unet &900 subjects & DSC = 0.554 \\
\hline
\end{longtable}


\subsection{BM Detection and Other BM Analysis Tasks}

\begin{figure}[h!]
\centering
\includegraphics[width=0.8\textwidth]{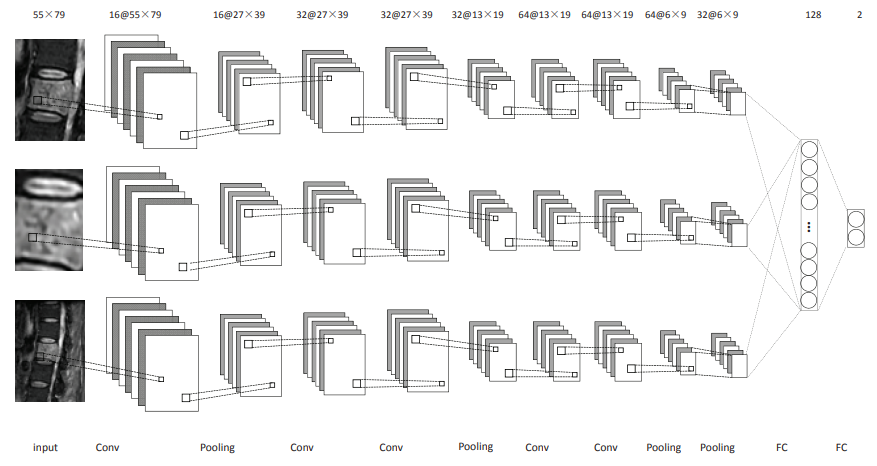}
\caption{ BM Detection System proposed by Wang et al.\cite{ wang2017multi}}
\label{wangetalarchi}
\end{figure}

In addition to classification and segmentation, numerous studies have focused on the development of novel approaches for various tasks related to metastatic bone lesions, including detection \cite{huo2023deep,kim2024automated}, prediction \cite{wakabayashi2021predictive} and registration\cite{yip2014use,noorda2014registration}. In the context of BM detection, Wang et al. \cite{wang2017multi} proposed a deep CNN architecture for detecting vertebral metastases in MRI scans. Their approach uses a Siamese Deep Neural Network (SdNN) in conjunction with multi-resolution analysis and weighted averaging of neighboring cross-sections. The objective of their approach is to exploit similarities and aggregate detection results. As shown in Figure \ref{wangetalarchi}), SdNN included three identical multilayer subnets to process image patches at different resolutions and generate likelihood maps for each MRI slice to enable final classification.

\begin{figure}[h!]
\centering
\includegraphics[width=0.8\textwidth]{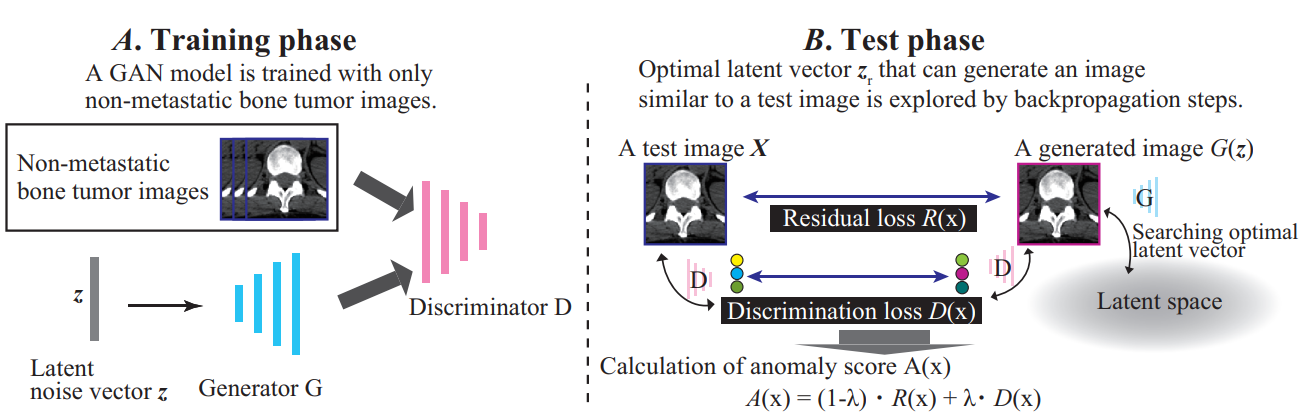}
\caption{ Overview of the based AnoGAN detection system proposed by Watanabe et al. \cite{watanabe2019bone}}
\label{watanabi}
\end{figure}

Similarly, Yao et al. \cite{Yao_2017,Yao_2012} proposed a detection system that used an SVM classifier \cite{cortes1995support} to identify different types of spinal lesions: sclerotic, lytic, and mixed. The first work in \cite{Yao_2012} achieved a sensitivity of 75\% for identifying sclerotic rib lesions. In their second work \cite{Yao_2017},their approach achieved sensitivities of 81\% for sclerotic and lytic lesions, while 76\% for mixed lesions. Recently, the authors in \cite{watanabe2019bone} proposed to use an abnormality detection approach for detecting BM in CT images. The proposed method is based on a generative adversarial network (AnoGAN) model. AnoGAN is a type of deep learning model consisting of a generator and a discriminator, designed to learn the manifold of normal anatomical variability in medical images. As can be seen in Figure \ref{watanabi}, the AnoGAN is trained only with images of non-metastatic bone tumors. Through unsupervised training, their proposed AnoGAN  is designed to generate images of non-metastatic tumors. When a CT test image is then fed into the model, an abnormality score is calculated to determine whether it contains a metastatic bone or not.
For the aim to automatically reduce false positive predictions of the detected hotspots in bone scintigraphy images, Providência et al. \cite{Provid_ncia_2021} proposed a semi-supervised  algorithm to the challenges associated with building a fully annotated database. Their approach achieved a classification sensitivity of 63\%, specificity of 58\%, and false-negative rate of 37\%.

Few studies have been presented in the area of BM registration, indicating a lack of research in this area. In a study by Yip et al. \cite{yip2014use} presented an automated method to match metastatic bone lesions using articulated registration, allowing for improving therapeutic assessment and reducing physician workload. Despite challenges such as varying contouring thresholds and segmentation imperfections, their approach outperformed rigid and deformable registration algorithms in BM lesion matching.
In another study by Noorda et al. \cite{noorda2014registration}, an automatic registration method was introduced for CT-MR registration of pre-treatment images in patients with bone metastases undergoing MR-HIFU treatment. Their method aims to register a single CT image on all available MRI images of the same patient.

For pain response prediction, Wakabayashi et al. \cite{wakabayashi2021predictive} carried out a retrospective investigation to predict pain response after radiotherapy for spinal metastases using a combination of clinical and radiomic characteristics. The study included data from 69 patients who received palliative radiation therapy, and the model that incorporated both types of features achieved the highest predictive performance, with an area under the curve (AUC) of 0.848. This research represents a groundbreaking effort to employ pretreatment CT radiomic features to predict pain response to radiation therapy in patients with spinal metastases, thus offering the potential for treatment approaches. 

\subsection{Multi-Tasks BM Analysis}

\begin{figure}[h!]
\centering
\includegraphics[width=0.8\textwidth]{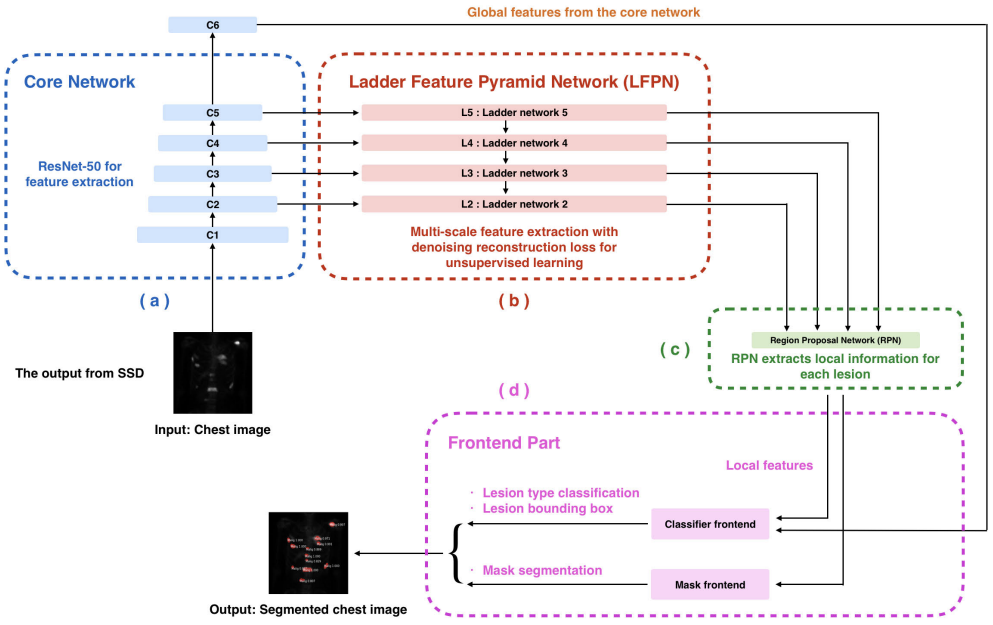}
\caption{ An overview of MaligNet model proposed by Apiparakoon et al.  \cite{apiparakoon2020malignet}}
\label{malignetarchi}
\end{figure}

\begin{figure}[h!]
\centering
\includegraphics[width=0.8\textwidth]{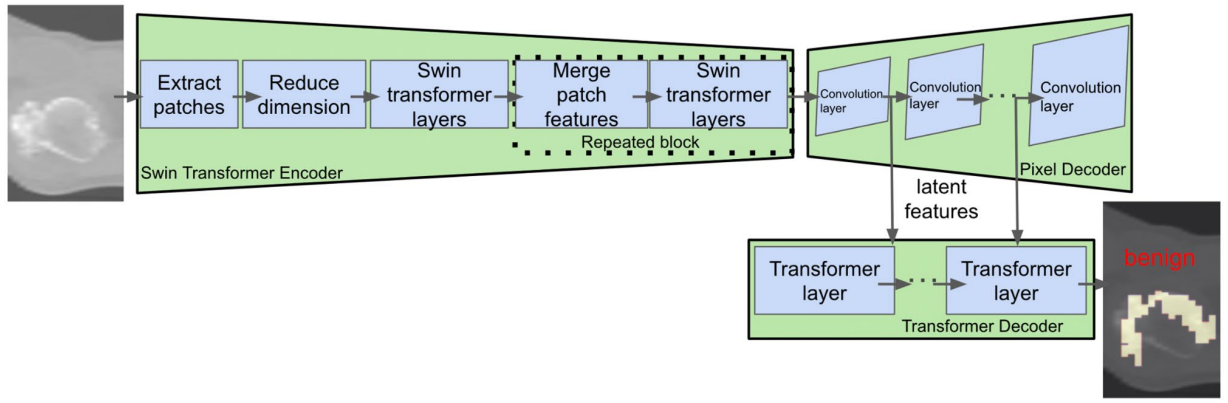}
\caption{ The proposed DL architecture of Yildiz et al. \cite{Yildiz_Potter_2023}.The architecture takes a DICOM slice from a CT volume as input and generates two outputs. Firstly, it predicts a segmentation mask that identifies the tumor region, which is visualized in yellow. Secondly, it provides a class prediction indicating whether the tumor is benign or malignant.}
\label{yildizarchi}
\end{figure}

In addition to the single BM analysis tasks discussed in the previous sections, Multi-Tasks have got increasing attention in the recent years. A notable study by Geng et al. \citep{Geng_2015} combined the tasks of BM classification and segmentation using a novel DL architecture. The researchers utilized the complementary information from both tasks to achieve better accuracy in the classification and delineation of BM lesions in the thoracic region. They used a sparse autoencoder \cite{raina2007self} and CNN to train an image-level classifier that categorized input images as normal or suspicious. For suspicious images, a patch-level classifier was trained using Multiple Instance Learning (MIL). Then an image-level classifier was trained to create a probability map of the hot spot. After acquiring the hotspot probability map, they use the LSD (Local Signed Difference) level set algorithm to delineate hotspots from the input image. The LSD algorithm is able to account for intensity in-homogeneity and weak object boundaries while taking into account global information, such as the arrangement of local clusters, resulting in robust segmentation performance \cite{wang2013region}. 
In \cite{apiparakoon2020malignet}, Apiparakoon et al. proposed a semi-supervised approach called MaligNet to segment bone lesion instances and to classify breast bone cancer metastases. The proposed MaligNet model is an instance segmentation model that incorporates ladder networks to utilize labeled and unlabeled data as shown in Figure \ref{malignetarchi}. Unlike traditional DL segmentation models, MaligNet utilizes global information through an additional connection from the core network.
A notable achievement of Yildiz et al. \cite{Yildiz_Potter_2023} was the integration of BM segmentation and classification into a comprehensive framework. In their retrospective study, an automated DL method for bone tumor segmentation and classification was developed using CT imaging (see Figure \ref{yildizarchi}). The study utilized a dataset of 84 femoral CT scans with final histologic confirmation of bone lesions and employed a DL architecture that predicted a segmentation mask over the estimated tumor region and a corresponding class as either benign or malignant. Despite an unbalanced dataset, the approach achieved comparable specificity (75\%) and sensitivity (79\%) and an average Dice score of 56\% for segmentation and up to 80\% for individual image slices.
Subsequently, in \cite{Cheng_20211}, the authors proposed an early diagnosis system for the identification and classification of BM in whole-body bone scan images. Their proposed approach uses a Deep Convolutional Neural Network (D-CNN) and shows satisfactory performance on a relatively small dataset of 205 cases, 100 of which have BM.The D-CNN achieved a sensitivity of 0.82 ± 0.08 and a precision of 0.70 ± 0.11 for the detection and classification of BM in the breast, while it achieved a sensitivity of 0.87 ± 0.12 and a specificity of 0.81 ± 0.11 for the classification of BM in the pelvis. The authors suggest the inclusion of hard example mining to improve the sensitivity and precision of breast D-CNN. Cheng et al.  \cite{cheng2021lesion} used the YOLOv4 model \cite{redmon2016you} to detect and classify BM in chest and pelvic scintigraphic images of patients with prostate and breast cancer. In their study, a novel approach was introduced using negative mining to prepare training samples for a different class to reduce the false positive rate. This innovative idea is based on the reasoning that it is more efficient to let the model identify the false positive patterns since they are not known. This procedure involved selecting negative cases for the model to test, and collecting the false positives for further training.

\begin{figure}[h!]
\centering
\includegraphics[width=0.8\textwidth]{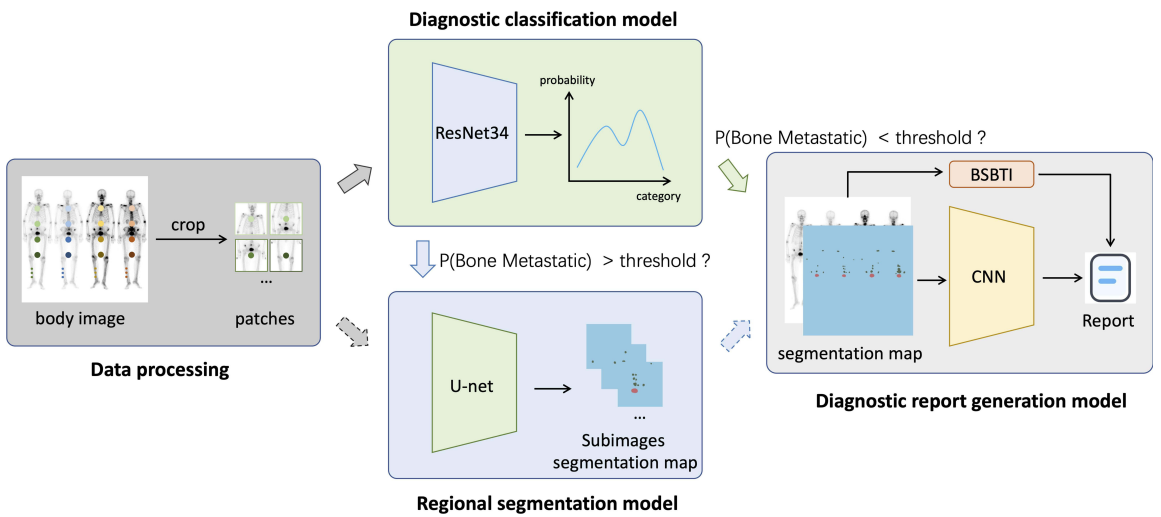}
\caption{ Overview of the prediction framework proposed by Liu et al.\citep{Liu_2022}}
\label{liuetalarchi}
\end{figure}

While previous works have attempted to solve multiple tasks simultaneously, such as segmentation and classification or detection and classification, the authors of \citep{Liu_2022} go one step further by proposing a system that mimics a radiologist. The methodology of this system includes several models (see Figure\ref{liuetalarchi}), including a classification model to identify BM, a segmentation model to automatically delineate the areas of the lesions, an evaluation model to quantify the tumor burden, and a model to generate diagnostic reports. The use of multitasking systems is very promising for BM detection. However, the practical clinical integration of such systems is a major challenge, as interoperability with existing medical infrastructure and rigorous validation processes are required to ensure reliability and regulatory compliance

\section{Discussion and Future Trends}
\label{sec8}
\begin{figure}[h!]
\centering
\includegraphics[width=0.8\textwidth]{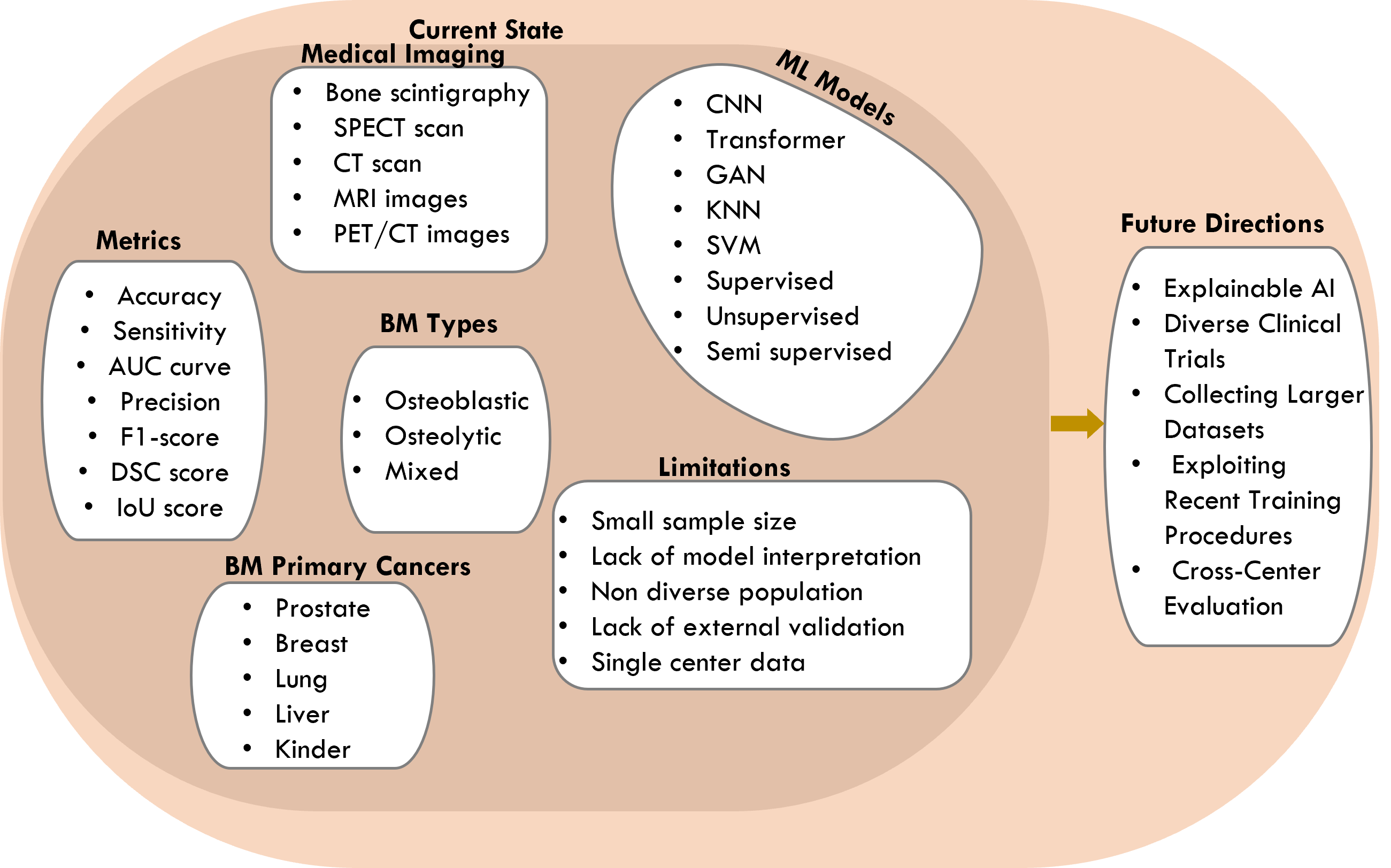}
\caption{ Artificial Intelligence in Metastatic Bone Research: A Comprehensive Overview and Prospective Directions}
\label{future}
\end{figure}

The applications of AI techniques in BM analysis have shown the potential for significant benefits and opportunities. While it is important to recognize AI-based methods in medical image analysis, the indispensable role of conventional methods, including manual examination by radiologists and clinicians, must also be acknowledged. These conventional methods rely on the years of experience and reliability of radiologists. Their value and expertise in visually detecting signs of BM and making accurate assessments must not be overlooked.
Figure \ref{future} shows a flowchart describing the different phases and components of a BM analysis system, while highlighting the importance of both AI techniques and traditional approaches for comprehensive analysis. This figure highlights the challenges prevalent in this area, including the limitations of AI models and the need for adequate data sets to ensure accurate and reliable results.

One of the most challenging aspects for BM analysis using AI, in particular DL, is the availability, the size and the quality of the annotated datasets. Currently, there are only a few datasets that provide comprehensive ground truth and accurate annotations. This is largely due to the high costs associated with obtaining and maintaining images. In addition, the existing datasets tend to focus on mass recognition, resulting in an unbalanced representation of related work. To overcome these challenges, it is crucial to develop  balanced and comprehensive datasets for training and evaluating AI models for BM analysis. In particular, achieving high performance in training DL models requires a large number of samples, although techniques such as semi-supervised or unsupervised have been extensively investigated to cope with this problem. However, the performance of these approaches still needs more investigation reduce the gap with the supervised procedure. In addition, providing accurate and systematic annotations is crucial, especially given the differences in tasks such as segmentation and classification. Some datasets use annotated regions of interest (ROIs), which are characterized by circles around the pixels of interest and may not be suitable as a basis for segmentation. To counteract subjectivity, the involvement of different radiologists in the annotation process can contribute to relatively objective annotations. Nevertheless, the limited availability of publicly available datasets poses a challenge when it comes to meeting all the requirements for training and evaluating DL models. To overcome the challenge of insufficient datasets, techniques such as transfer learning and data augmentation have been widely exploited. Generally, the state-of-the art approaches use pretrained weights from non-medical domain to initialize their models weights before starting the training. Although the usefulness of this approach, research on knowledge transfer between different medical image modalities or BM datasets needs to be highlighted in the future researches, which has a high potential to provide new direction to achieve more precise results. On the other hand, only traditional data augmentation techniques have been used. Recent data augmentation techniques such as cut-mix and super-pixel can be investigated to achieve better generalization performance.

Deep learning-based systems generally outperform conventional methods. However, the overall performance of modern computer-aided bone metastasis detection (CAD) systems remains unsatisfactory, especially in detecting small lesions, which presents a particular challenge. Incorporating deep learning expertise can enhance interpretability and provide better understanding behind convolutional operations. The interpretability of deep learning models has historically been a concern, as they are often perceived as black boxes lacking the meaningful hand-crafted features present in traditional CAD systems. Efforts have been made to address this issue through deep learning visualization techniques.
Advanced DL models, such as attention-based models, hold promise for improving future CAD systems for bone metastases. Merging the advanced deep learning approaches with the recent developments on training paradigms like self-supervised learning, semi-supervised learning, and active learning can provide the right solution for moving towards real-world applications for BM analysis. Future research on CAD systems for bone metastases is likely to integrate these frameworks to enable better interpretation of models and their decision-making processes.

\section{Conclusion}
\label{sec9}

This survey provides a comprehensive overview of recent work on the application of AI for BM analysis, focusing on the works from the last 13 years. The focus is on ML, particularly DL, for purposes such as diagnosis, decision support, treatment support and prognosis in bone cancer. Furthermore, the review highlights applications in the areas of classification, detection, segmentation and registration.

ML technologies, especially DL, show promising performance in improving clinician efficiency and reducing adverse events in the treatment of bone cancer. The versatility of ML in various clinical aspects underlines its potential. On the other hand, the survey highlights the need for rigorous validation studies to ensure reliability and generalizability in different clinical settings. Seamless integration of ML into routine clinical practice is crucial.

Collaboration between clinicians, researchers and technology developers is essential to realize the full potential of ML in the treatment of BM. This review highlights the importance of generating more datasets for comparison between different ML approaches. The lack of publicly available datasets is hampering progress in this field. Moreover, advanced ML approaches such as unsupervised learning, semi-supervised learning, data distillation and self-supervised learning as well as newer architectures such as Transformers and GAN need to be further explored for BM analysis.



\end{document}